\documentclass[12pt, a4paper]{article}
\usepackage{geometry}
\usepackage{amstext}
\usepackage{amsmath}
\usepackage{graphicx}
\usepackage{subfigure}
\usepackage{mathrsfs}
\usepackage{indentfirst}
\usepackage{amssymb}
\usepackage{color}
\usepackage{cite}
\usepackage{multirow}
\usepackage[utf8]{inputenc}
\title{Determinations of form factors for semileptonic $D\rightarrow K$ decays and leptoquark constraints}

\author{Jian Zhang\thanks{E-mail:zhangjianphy@aliyun.com} \and Chong-Xing Yue\thanks{E-mail:cxyue@lnnu.edu.cn} \and Chun-Hua Li\thanks{E-mail:chunhua@lnnu.edu.cn}\\
	{\small Department of Physics, Liaoning  Normal University, Dalian
		116029, P. R. China}
}
\date{\today}

\begin{document}
	\maketitle
	\begin{abstract}
		By analyzing all existing measurements 
		for $ D\rightarrow K \ell^+ \nu_{\ell} $ ( $\ell=e,\ \mu$ ) 
		decays, we find that the determinations of 
		both the vector form factor $f_+^K(q^2)$ and 
		scalar form factor $f_0^K(q^2)$ for 
		semileptonic $D\rightarrow K$ decays from 
		these measurements are feasible. 
		By taking the parameterization of the one order series expansion
		of the $f_+^K(q^2)$ and $f_0^K(q^2)$, $f_+^K(0)|V_{cs}|$ is
		determined to be $0.7182\pm0.0029$, and the shape parameters of 
		$f_+^K(q^2)$ and $f_0^K(q^2)$ are $r_{+1}=-2.16\pm0.007$ and 
		$r_{01}=0.89\pm3.27$, respectively.  
		Combining with the average $f_+^K(0)$ of $N_f=2+1$ and $N_f=2+1+1$
		lattice calculaltion, the $|V_{cs}|$ is extracted 
		to be $0.964\pm0.004\pm0.019$ where the first error is experimental 
		and the second theoretical. 
		Alternatively, the $f_+^K(0)$ is extracted to be $0.7377\pm0.003\pm0.000$ by taking 
		the $|V_{cs}|$ as the value from the global fit with the unitarity constraint of 
		the CKM matrix.
		Moreover, using the obtained form factors 
		by $N_f=2+1+1$ lattice QCD, we re-analyze these 
		measurements in the context of new physics. Constraints 
		on scalar leptoquarks are obtained for different final 
		states of semileptonic $D \rightarrow K$ decays.
	\end{abstract}

\section{Introduction}
\label{intro}
Semileptonic $D\rightarrow P(P=K, \pi)$ decays have 
long been of great interest in the field of flavor physics.
They play important roles in validating the 
lattice QCD (LQCD), extracting 
the Cabibbo-Kobayashi-Maskawa (CKM) matrix elements, 
and searching for New Physics (NP) beyond the Standard Model (SM) \cite{D-review}.

For the decay $ D\rightarrow K \ell^+ \nu_{\ell} $ ( $\ell=e,\ \mu$ ), 
strong and weak interaction portions can be well separated 
and the effects of strong interactions can be 
parameterized by form factors. 
In the SM, the differential decay rate as a function of $q^2$ is given by
\begin{eqnarray}
\label{eq:DKSM}
\dfrac{\mathrm{d}\Gamma(D\rightarrow K \ell^+ \nu_{\ell})}{\mathrm{d}q^{2}} &=& \frac{G_{F}^{2} \rvert V_{cs} \rvert^{2}}{24\pi^{3}} \rvert \mathrm{\textbf{p}} \rvert^3 \left(1-\frac{m_{\ell}^2}{q^2} \right)^2 
\nonumber
\\
& \cdot&  \left\{ \left( 1+\dfrac{m_{\ell}^{2}}{2q^{2}} \right) \rvert f_{+}^K(q^{2}) \rvert^{2}+  \dfrac{3m_{\ell}^{2} (m_{D}^{2}-m_{K}^{2})^{2}}{8m_D^2\rvert \mathrm{\textbf{p}} \rvert^2q^{2}} \rvert f_{0}^K(q^{2}) \rvert^{2} \right\},
\end{eqnarray}
where $G_{F}$ is the Fermi constant, $\textbf{p}$ represents 
the three momentum of the $K$ meson in the $D$ rest frame, 
and $q\equiv p_D - p_K$ is the four momenta transferred 
to $ \ell^+ \nu_{\ell}$ pair.
The range of $q^2$ is from $m_\ell^2$ 
when $K$ has the maximum possible momentum 
to $(m_{D}-m_{K})^{2}$ when the $K$ meson is at rest. 
The vector form factor $f_{+}^K(q^{2})$ and the scalar form factor $f_{0}^K(q^{2})$ are defined via
\begin{eqnarray}
 \langle K(p_K)|\bar{s}\gamma^\mu c|D(p_D)\rangle  =\left(p_D^\mu+p_K^\mu-\dfrac{m_D^2-m_K^2}{q^2}q^\mu \right)f_{+}^K(q^{2})+ \dfrac{m_D^2-m_K^2}{q^2}q^\mu f_{0}^K(q^{2}) ,
\end{eqnarray} 
and
\begin{eqnarray}
\langle K(p_K)|\bar{s} c|D(p_D)\rangle=\dfrac{m_D^2-m_K^2}{m_c-m_s} f_{0}^K(q^{2}).
\end{eqnarray} 
At the maximal recoil point, kinematic constraints lead $f_{+}^K(0)=f_{0}^K(0)$. 

In the last 30 years, various measurements 
of the decay $D\rightarrow K \ell^+ \nu_{\ell}$ were performed at 
more than ten experiments. 
The decay rates of $D^{0}\rightarrow K^{-}\ell^{+}\nu_{\ell}$
and $D^{+}\rightarrow \bar{K}^{0}\ell^{+}\nu_{\ell}$ in different $q^2$ bins 
were measured at the experiments the E691 \cite{E691}, E687\cite{E6871993,E6871995}, E653\cite{E653},
Mark-III \cite{Mark-III-1989},CLEO \cite{CLEO-1991}, FOCUS\cite{FOCUS20041},
CLEO-II \cite{CLEO-II-1993}, BaBar \cite{BaBar-2007}, 
BES-II \cite{BES-II-2004,BES-II-2005,BESII20061,BESII2006},CLEO-c \cite{CLEO-c-2009} 
and BES-III \cite{BES-III-2015,BES-III-2016,BES-III-2016mu,BES-III-2017}. 
The FOCUS experiment measured non-parametric relative 
form factor from $D^0\rightarrow K^- \mu^+ \nu_{\mu}$ in 2005 \cite{FOCUS-2004}, 
and the Belle experiment measured 
the vector form factor from $D^0\rightarrow K^- \ell^+ \nu_{\ell}$ 
in 2006 \cite{Belle2006}. By combining these measurements, one can obtain $f_+^K(0)\rvert V_{cs} \rvert$, 
the product of the hadronic form factor at $q^2=0$
and the magnitude of CKM matrix 
element $V_{cs}$.
With the values of $ \rvert V_{cs} \rvert$ from the the global fit
with the unitarity constraint of CKM matrix 
and $f_+^K(0)$ calculated in lattice QCD, $f_+^K(0)$ and 
$ \rvert V_{cs} \rvert$ can be extracted from $f_+^K(0)\rvert V_{cs} \rvert$,
respectively \cite{Fang2014}.
In 2014, ref. \cite{Fang2014} extract $f_{+}^K(0)$ and $\rvert V_{cs} \rvert$ by 
considering all the experimental measurements 
of $D\rightarrow K e^+ \nu_{e}$ decays before 2014. 

In these experimental and theoretical studies, the contribution 
of $f_{0}$ term is neglected since it is suppressed by the mass squared of lepton.
However, with the improvement of 
experimental precision, it is feasible to determine  
both the vector and scalar form factors. 
Here, we determine both $f_{+}^K(q^{2})$ and $f_{0}^K(q^{2})$ 
for the first time by comprehensively analyzing all the 
experimental measurements of $D\rightarrow K \ell^{+}\nu_{\ell}$. 
As the result of this analysis, we report the values of 
$f_+^K(0)\rvert V_{cs} \rvert$, $r_{+1}$ and $r_{01}$ which are
the shape parameters of $f_{+}^K(q^{2})$ and $f_{0}^K(q^{2})$, respectively. 
We determine $f_+^K(0)$ from $f_+^K(0)\rvert V_{cs} \rvert$ by taking $ \rvert V_{cs} \rvert$
as the value obtained from the global fit with the unitarity constraint 
of CKM matrix done by the Particle Data Group (PDG) in 2016 \cite{globalfit}. 
$ \rvert V_{cs} \rvert$ is extracted with the value of $f_+^K(0)$ 
calculated in LQCD.

In addition, a comprehensive analysis of these measurements 
is important to search for  non-Standard interactions beyond the Standard weak 
to $D\rightarrow K \ell^+ \nu_{\ell}$. One candidate of the 
non-Standard interactions is to exchange a scalar 
leptoquark \cite{Kronfeld2008,Dobrescu2008,Barranco2016}. 
Leptoquarks are hypothetical color-triplet bosons that carry both 
baryon number and lepton number, and can thus couple directly 
to a quark and a lepton \cite{BRW1986,LQ-review}. Leptoquark can be of either vector (spin-1)
or scalar (spin-0) nature according to their properties under the 
Lorentz transformations. Some scalar leptoquarks can lead to the 
effective $\bar{s}c\bar{\nu}\ell$ vertex. Searching for the scalar leptoquarks 
from $D\rightarrow K \ell^{+}\nu_{\ell}$ is one of 
the goals of this article. By taking the form factors from lattice calculations, 
we re-analyze the experimental measurements of $ D\rightarrow K \ell \nu_{\ell} $ 
in the context of new physics and provide the 
constraint on scalar leptoquark. 

The article is organized as follows: We review the parameterization of the form factors in Section \ref{sec:2} firstly,
and then present the details of the experimental measurements 
of $D\rightarrow K \ell^{+}\nu_{\ell}$ in Section \ref{sec:3}. 
The procedure of the analysis is described in Section \ref{sec:4}.
In Section \ref{sec:5}, 
we study these experimental measurements in the context of new physics.
Finally, the conclusions of this work are given in Section \ref{sec:6}.

\section{Parameterization of the form factors}
\label{sec:2}
The form factors $f_+^K(q^2)$ and $f_0^K(q^2)$ 
can be parameterized according to the constraints 
of their general properties of analyticity, 
cross symmetry, and unitarity\cite{1008.1857}. 
Various parameterizations 
exist such as the single 
pole model\cite{singlepole}, the modified 
pole model\cite{singlepole}, the $ISGW2$ model\cite{ISGW2} 
and the $series \ expansion$\cite{seriesexpansion}. 
The experimental data, however, does not 
support the former three models well \cite{Fang2014}, so the $series \ expansion$ 
is used in this article. 
In this parameterization, the form factors transformed from $q^2$-space to $z$-space, where 
\begin{equation}
z(q^2,t_0)=\dfrac{\sqrt{t_+ - q^2} - \sqrt{t_+ - t_0}}{\sqrt{t_+ - q^2} + \sqrt{t_+ - t_0}},
\end{equation}
with $t_\pm = (m_D \pm m_K)^2$ and $t_0 = t_+ (1 - \sqrt{1-t_- / t_+} )$. The form factors is then expressed as

\begin{equation}
\label{form factor}
f_{+(0)}^K(q^2) = \dfrac{1}{\mathcal{P}_{+(0)}(q^2) \phi(q^2, t_0)} \sum_{k=0}^\infty a_k^{+(0)}(t_0)[z(q^2, t_0)]^k,
\end{equation}
where $a_k(t_0)$ are real coefficients. The function $\mathcal{P}_+(q^2)$ is $\mathcal{P}_+(q^2) = z(q^2, m_{D_s^*}^2)$ for $f_+^K(q^2)$ and $\mathcal{P}_0(q^2)$ is $\mathcal{P}_0(q^2) = z(q^2, m_{D_{s0}^*}^2)$ for $f_0^K(q^2)$. $ \phi(q^2, t_0) $ is chosen to be
\begin{eqnarray}
\phi(q^2,t_0) = \left( \frac{\pi m^2_c}{3} \right)^{1/2} \left( \frac{z(q^2,0)}{-q^2} \right)^{5/2} \left( \frac{z(q^2,t_0)}{t_0-q^2} \right)^{-1/2} \left( \frac{z(q^2,t_-)}{t_--q^2} \right)^{-3/4} \frac{(t_+-q^2)}{(t_+-t_0)^{1/4}},
\end{eqnarray}
where $m_c$ is the mass of the charm quark.

By using the relation $1 = f_+^K(0)\mathcal{P}(0) \phi(0, t_0)/ ( \sum_{k=0}^\infty a_k(t_0)[z(0, t_0)]^k ) $ 
deduced from Eq.(\ref{form factor}), we obtain
\begin{equation}
\label{form factor 2}
f^K(q^2) = \dfrac{f_+^K(0)\mathcal{P}(0) \phi(0, t_0)(1 + \sum_{k=1}^Nr_k[z(q^2, t_0)]^k)}{\mathcal{P}(q^2) \phi(q^2, t_0)(1 + \sum_{k=1}^Nr_k[z(0, t_0)]^k)},
\end{equation}
where $r_{k}=a_k(t_0)/a_0(t_0)$ and $N$ is the expansion order. 

\section{Experimental measurements }
\label{sec:3}
\begin{table}
	\caption{Ratios $\mathcal{R}_{0(+)}^{\ell}$ measured at different experiments.}
	\label{tabrDD}
	\centering
	\begin{tabular}{lll}
		\hline
		\hline\noalign{\smallskip}
		Experiment         & $q^2$ (GeV) & $\mathcal{R}_{0(+)}^{\ell}$  \\
		\noalign{\smallskip}\hline\noalign{\smallskip}
		E691\cite{E691}    & ($m_{e}^2,\ q_{max}^2$) &$\mathcal{R}_{0}^{e}$ = 0.91 $\pm$ 0.07 $\pm$ 0.11 \\
		E687\cite{E6871993}   & ($m_{\mu}^2,\ q_{max}^2$) &$\mathcal{R}_{0}^{\mu}$ = 0.82 $\pm$ 0.13 $\pm$ 0.13 \\
		E687\cite{E6871995}   & ($m_{\mu}^2,\ q_{max}^2$) &$\mathcal{R}_{0}^{\mu}$ = 0.852 $\pm$ 0.034 $\pm$ 0.028 \\
		CLEO\cite{CLEO-1991}    & ($m_{e}^2,\ q_{max}^2$) &$\mathcal{R}_{0}^e$ = 0.90 $\pm$ 0.06 $\pm$ 0.06 \\
		CLEO\cite{CLEO-1991} & ($m_{\mu}^2,\ q_{max}^2$) & $\mathcal{R}_0^{\mu}= 0.79 \pm 0.08 \pm 0.09$ \\
		CLEO-II\cite{CLEO-II-1993} & ($m_{e}^2,\ q_{max}^2$) & $\mathcal{R}_{0}^e$ = 0.978 $\pm$ 0.027 $\pm$ 0.044\\
		CLEO-II\cite{CLEO-II-1993} & ($m_{e}^2,\ q_{max}^2$) & $\mathcal{R}_{+}^e$ = 2.60 $\pm$ 0.35 $\pm$ 0.26\\ 
		BaBar\cite{BaBar-2007} 
		& ($m_{e}^2,\ q_{max}^2$) & $\mathcal{R}_{0}^e$ = 0.927 $\pm$ 0.007 $\pm$ 0.012\\ 	
		\noalign{\smallskip}\hline
		\hline
	\end{tabular}
\end{table}

\begin{table}
	\caption{Branching fractions $\mathcal{B}_{0(+)}^{\ell}(D\rightarrow K \ell^+ \nu_{\ell})$ measured at different experiments.}
	\label{tabbDD}
	\centering
	\begin{tabular}{lll}
		\hline
		\hline\noalign{\smallskip}
		Experiment         & $q^2$ (GeV) & $\mathcal{B}_{0(+)}^{\ell} (\%)$  \\
		\noalign{\smallskip}\hline\noalign{\smallskip}
		E653\cite{E653} & ($m_{\mu}^2,\ q_{max}^2$) &$\mathcal{B}_0^{\mu}= 3.16 \pm 0.52$\\
		Mark-III\cite{Mark-III-1989} & ($m_{e}^2,\ q_{max}^2$) &$\mathcal{B}_0^e= 3.4 \pm 0.5 \pm 0.4$\\
		FOCUS\cite{FOCUS20041} & ($m_{\mu}^2,\ q_{max}^2$) &$\mathcal{B}_+^{\mu}= 9.15 \pm 0.94$\\
		BES-II\cite{BES-II-2004}    & ($m_{e}^2,\ q_{max}^2$) & $\mathcal{B}_0^e= 3.82 \pm 0.40 \pm 0.27$ \\
		BES-II\cite{BES-II-2005}    & ($m_{e}^2,\ q_{max}^2$) &$\mathcal{B}_+^e = 8.95 \pm 1.59 \pm 0.67$\\
		BES-II \cite{BESII20061}  & ($m_{\mu}^2,\ q_{max}^2$) &$\mathcal{B}_0^{\mu} = 3.55 \pm 0.56 \pm 0.59$\\
		BES-II\cite{BESII2006}    & ($m_{\mu}^2,\ q_{max}^2$) &$\mathcal{B}_+^{\mu} = 10.3 \pm 2.3 \pm 0.8$\\
		BES-III\cite{BES-III-2016}    & ($m_{e}^2,\ q_{max}^2$) &$\mathcal{B}_+^e = 8.59 \pm 0.14 \pm 0.21$\\
		BES-III\cite{BES-III-2016mu}    & ($m_{\mu}^2,\ q_{max}^2$) &$\mathcal{B}_+^{\mu} = 8.72 \pm 0.07 \pm 0.18$\\
		\noalign{\smallskip}\hline
		\hline
	\end{tabular}
\end{table}

The existing measurements for $D^{0}\rightarrow K^{-}\ell^{+}\nu_{\ell}$ and $D^{+}\rightarrow \bar{K}^{0}\ell^{+}\nu_{\ell}$ can be divided into three categories:
\\

(i) Ratio of the branching fractions $\mathcal{R}_{0(+)}^{\ell}$, where $\mathcal{R}_0^{\ell}=\mathcal{B}^{\ell}(D^{0}\rightarrow K^{-}\ell^{+}\nu_{\ell})/\mathcal{B}(D^{0}\rightarrow K^{-}\pi^{+})$, and $\mathcal{R}_+^{\ell}=\mathcal{B}^{\ell}(D^{+}\rightarrow \bar{K}^{0}\ell^{+}\nu_{\ell})/\mathcal{B}(D^{+}\rightarrow \bar{K}^{0}\pi^{+})$. The ratios measured at different experiments are listed in Table \ref{tabrDD}. 

(ii) Decay branching fraction $\mathcal{B}_{0(+)}^{\ell}$, 
where $\mathcal{B}_{0}^{\ell}$ and $\mathcal{B}_{+}^{\ell}$ is the branching fraction 
of $D^{0}\rightarrow K^{-}\ell^{+}\nu_{\ell}$ and $D^{+}\rightarrow \bar{K}^{0}\ell^{+}\nu_{\ell}$, respectively. 
The measurements of $\mathcal{B}_{0(+)}^{\ell}$ at different experiments are shown in Tab. \ref{tabbDD}.  For the sake of convenience, the radios $\Gamma(D^0\rightarrow K^-\mu^+\nu_{\mu})/\Gamma(D^0\rightarrow \mu^+X)$ = 0.472 $\pm$ 0.051 $\pm$ 0.040 measured at E653 experiment \cite{E653} and $\Gamma(D^+\rightarrow \bar{K}^0\mu^+\nu_{\mu})/\Gamma(D^+\rightarrow K^-\pi^+\pi^+)$ = 1.019 $\pm$ 0.076 $\pm$ 0.065 measured at the FOCUS experiment \cite{FOCUS20041} have been transformed into corresponding branching fractions also listed in Tab. \ref{tabbDD} by using $B(D^0\rightarrow \mu^+X)=(6.7 \pm 0.6)$\% and $B(D^+\rightarrow K^-\pi^+\pi^+)=(8.98 \pm 0.28)$\% which are taken from PDG \cite{PDG}.

(iii) Decay rate $\Delta\Gamma$, where $\Delta\Gamma$ represents the partial decay 
rate of $D^{0}\rightarrow K^{-}e^{+}\nu_{e}$ 
or $D^{+}\rightarrow \bar{K}^{0}e^{+}\nu_{e}$ in a certain $q^2$ bin. 
\\

Measurements of the first two categories could not 
be used directly to determine $f_+^K(0)\arrowvert V_{cs}\arrowvert$ 
and the shapes of form factors. 
To use these measurements, we should 
first transfer them into absolute decay rates in certain $q^2$ ranges \cite{Fang2014}.

The absolute decay rates for the experimental results classified as
the categories (i) and (ii) measurements can be extracted respectively by
\begin{equation}
\Delta\Gamma = \mathcal{R} \times \mathcal{B}(D\rightarrow K\pi) \times \dfrac{1}{\tau_{D}},
\end{equation}
and 
\begin{equation}
\Delta\Gamma =\mathcal{B}(D\rightarrow K\ell^{+}\nu_{\ell}) \times \dfrac{1}{\tau_{D}},
\end{equation}
where $\mathcal{B}(D\rightarrow K\pi)$ is the branching fraction 
for $D^{0}\rightarrow K^{-}\pi^{+}$ 
or $D^{+}\rightarrow \bar{K}^{0}\pi^{+}$ decays, 
and $\tau_{D}$ is the lifetime of $D^{0}$ or $D^{+}$ meson. 
To avoid the possible correlations, 
we use $\mathcal{B}(D^{0}\rightarrow K^{-}\pi)= (3.89 \pm 0.04)\%$, $\mathcal{B}(D^{+}\rightarrow \bar{K}^{-}\pi^{+})= (2.93 \pm 0.094)\%$  
which is the sum of $\mathcal{B}(D^{+}\rightarrow K_{S}^{0}\pi^{+})= (1.47 \pm 0.08)\%$ 
and $\mathcal{B}(D^{+}\rightarrow K_{L}^{0}\pi^{+})= (1.46 \pm 0.05)\%$), $\tau_{D^{0}}= (410.1 \pm 1.5) \times 10^{-15}$ s, $\tau_{D^{+}}=(1040 \pm 7)\times 10^{-15}$ s from PDG\cite{PDG}.

The absolute decay rates after the transformations and the measurements,
classified as the category (iii), of partial decay rates 
in different $q^2$ bins for $D\rightarrow K e^+ \nu_{e}$, are shown in Tabs. \ref{tabD0} and \ref{tabD1}.

\begin{table}
	\caption{Partial decay rates $\Delta\Gamma$ of $D^{0}\rightarrow K^{-}\ell^{+}\nu_{\ell}$ decays in $q^2$ ranges.}
	\label{tabD0}
	\centering
	\begin{tabular}{llr}
		\hline
		\hline\noalign{\smallskip}
		Experiment         & $q^2$ (GeV) & $\Delta\Gamma $ (ns$^{-1}$) \\
		\hline
		E691\cite{E691}    & ($m_{e}^2,\ q_{max}^2$) & 86.32 $\pm$ 12.40 \\
		CLEO\cite{CLEO-1991}    & ($m_{e}^2,\ q_{max}^2$) & 85.37 $\pm$ 8.10 \\
		CLEO-II\cite{CLEO-II-1993} & ($m_{e}^2,\ q_{max}^2$) & 92.77 $\pm$ 5.00 \\ 
		BaBar\cite{BaBar-2007}
		&($m_{e}^2,\ q_{max}^2$) & 87.92 $\pm$ 1.63 \\
		E687\cite{E6871993}   & ($m_{\mu}^2,\ q_{max}^2$) & 77.78 $\pm$ 17.46 \\
		E687\cite{E6871995}   & ($m_{\mu}^2,\ q_{max}^2$) & 80.82 $\pm$ 4.27 \\
		CLEO\cite{CLEO-1991}    & ($m_{\mu}^2,\ q_{max}^2$) & 74.94 $\pm$ 11.45 \\
		\hline
		Mark-III\cite{Mark-III-1989} & ($m_{e}^2,\ q_{max}^2$) & 82.91 $\pm$ 15.62 \\
		BES-II\cite{BES-II-2004}    & ($m_{e}^2,\ q_{max}^2$) & 93.15 $\pm$ 11.77 \\
		E653\cite{E653}   & ($m_{\mu}^2,\ q_{max}^2$) & 77.11 $\pm$ 12.65 \\
		BES-II\cite{BESII20061}     & ($m_{\mu}^2,\ q_{max}^2$) & 86.56 $\pm$ 19.84 \\
		\noalign{\smallskip}\hline\noalign{\smallskip}
		\multirow{1}{*}{CLEO-c\cite{CLEO-c-2009}}
		&($m_{e}^2$,\ 0.2) & 17.82 $\pm$ 0.43 \\
		&(0.2,\ 0.4) & 15.83 $\pm$ 0.39 \\	
		&(0.4,\ 0.6) & 13.91 $\pm$ 0.36 \\
		&(0.6,\ 0.8) & 11.69 $\pm$ 0.32 \\   
		&(0.8,\ 1.0) & 9.36  $\pm$ 0.28 \\
		&(1.0,\ 1.2) & 7.08  $\pm$ 0.24 \\	
		&(1.2,\ 1.4) & 5.34  $\pm$ 0.21 \\
		&(1.4,\ 1.6) & 3.09  $\pm$ 0.16 \\  
		&(1.6,\ $q_{max}^2$) & 1.28 $\pm$ 0.11 \\    
		\noalign{\smallskip}\hline\noalign{\smallskip}
		\multirow{1}{*}{BES-III\cite{BES-III-2015}}
		&($m_{e}^2$,\ 0.1) & 8.812 $\pm$ 0.187 \\
		&(0.1,\ 0.2) & 8.743 $\pm$ 0.162 \\
		&(0.2,\ 0.3) & 8.295 $\pm$ 0.159 \\	
		&(0.3,\ 0.4) & 7.567 $\pm$ 0.153 \\	
		&(0.4,\ 0.5) & 7.486 $\pm$ 0.152 \\
		&(0.5,\ 0.6) & 6.446 $\pm$ 0.138 \\
		&(0.6,\ 0.7) & 6.200 $\pm$ 0.134 \\   
		&(0.7,\ 0.8) & 5.519 $\pm$ 0.126 \\ 
		&(0.8,\ 0.9) & 5.028 $\pm$ 0.119 \\
		&(0.9,\ 1.0) & 4.525 $\pm$ 0.111 \\
		&(1.0,\ 1.1) & 3.972 $\pm$ 0.103 \\	
		&(1.1,\ 1.2) & 3.326 $\pm$ 0.093 \\
		&(1.2,\ 1.3) & 2.828 $\pm$ 0.085 \\
		&(1.3,\ 1.4) & 2.288 $\pm$ 0.077 \\
		&(1.4,\ 1.5) & 1.737 $\pm$ 0.068 \\ 
		&(1.5,\ 1.6) & 1.314 $\pm$ 0.058 \\  
		&(1.6,\ 1.7) & 0.858 $\pm$ 0.050 \\  
		&(1.7,\ $q_{max}^2$) & 0.379 $\pm$ 0.039 \\    
		\noalign{\smallskip}\hline
		\hline
	\end{tabular}
\end{table}

\begin{table}
	\caption{Partial decay rates $\Delta\Gamma$ of $D^{+}\rightarrow \bar{K}^{-}\ell^{+}\nu_{\ell}$ decays in $q^2$ ranges.}
	\label{tabD1}
	\centering
	\begin{tabular}{llr}
		\hline
		\hline\noalign{\smallskip}
		Experiment         & $q^2$ (GeV) & $\Delta\Gamma $ (ns$^{-1}$) \\
		\hline
		CLEO-II\cite{CLEO-II-1993} & ($m_{e}^2,\ q_{max}^2$) & 73.25 $\pm$ 12.52 \\
		BES-II\cite{BES-II-2005} & ($m_{e}^2,\ q_{max}^2$) & 86.06 $\pm$ 16.60 \\
		BES-III\cite{BES-III-2016} & ($m_{e}^2,\ q_{max}^2$) & 82.60 $\pm$ 2.49 \\
		FOCUS\cite{FOCUS20041} & ($m_{\mu}^2,\ q_{max}^2$) & 87.99 $\pm$ 9.08 \\
		BES-II\cite{BESII2006}    & ($m_{\mu}^2,\ q_{max}^2$) &$99.04 \pm 23.42$\\
		BES-III\cite{BES-III-2016mu}    & ($m_{\mu}^2,\ q_{max}^2$) &$83.85\pm 1.94$\\
		\noalign{\smallskip}\hline\noalign{\smallskip}
		\multirow{1}{*}{CLEO-c\cite{CLEO-c-2009}}
		&($m_{e}^2$,\ 0.2) & 17.79 $\pm$ 0.65 \\
		&(0.2,\ 0.4) & 15.62 $\pm$ 0.59 \\	
		&(0.4,\ 0.6) & 14.02 $\pm$ 0.54 \\
		&(0.6,\ 0.8) & 12.28 $\pm$ 0.49 \\   
		&(0.8,\ 1.0) & 8.92  $\pm$ 0.41 \\
		&(1.0,\ 1.2) & 8.17  $\pm$ 0.37 \\	
		&(1.2,\ 1.4) & 4.96  $\pm$ 0.27 \\
		&(1.4,\ 1.6) & 2.67  $\pm$ 0.19 \\  
		&(1.6,\ $q_{max}^2$) & 1.19 $\pm$ 0.13 \\ 
		\noalign{\smallskip}\hline\noalign{\smallskip}
		\multirow{1}{*}{BES-III\cite{BES-III-2017}}
		&($m_{e}^2$,\ 0.2) & 16.97 $\pm$ 0.60 \\
		&(0.2,\ 0.4) & 15.29 $\pm$ 0.53 \\	
		&(0.4,\ 0.6) & 13.57 $\pm$ 0.47 \\
		&(0.6,\ 0.8) & 11.65 $\pm$ 0.40 \\   
		&(0.8,\ 1.0) & 9.33  $\pm$ 0.34 \\
		&(1.0,\ 1.2) & 7.06  $\pm$ 0.28 \\	
		&(1.2,\ 1.4) & 4.96  $\pm$ 0.20 \\
		&(1.4,\ 1.6) & 2.97  $\pm$ 0.14 \\  
		&(1.6,\ $q_{max}^2$) & 1.01 $\pm$ 0.07 \\      
		\noalign{\smallskip}\hline
		\hline
	\end{tabular}
\end{table}

We also consider the non-parametric relative 
form factors $\mathfrak{f}_+^K(q^2)$ 
for $D^{0}\rightarrow K^{-}\mu^{+}\nu_{\mu}$ 
measured at the FOCUS experiment in 2005 \cite{FOCUS-2004}. 
The average values of relative form factors $\mathfrak{f}_+^K(q^2)$ in nine $q^2$ bins were obtained 
by assuming $f_+^K(0)$ has been normalized to 1 and 
the ratio $f_-^K(q^2)/f_+^K(q^2)= -0.7$, 
where $f_-^K(q^2)=(f_0^K(q^2)-f_+^K(q^2))(m_D^2-m_K^2)/q^2$. 
The measurements are listed in Tab. \ref{tabFOC}.

\begin{table}
	\caption{Non-parametric relative form factors $\mathfrak{f}_+^K(q_i^2)$ measured at the FOCUS experiment.}
	\label{tabFOC}
	\centering
	\begin{tabular}{lcc}
		\hline
		\hline\noalign{\smallskip}
		$i $        & $q^2$ (GeV) & $\mathfrak{f}_+^K(q^2)$ \\
		\noalign{\smallskip}\hline\noalign{\smallskip}
		1 & 0.09 & 1.01 $\pm$ 0.03 \\
		2 & 0.27 & 1.11 $\pm$ 0.05 \\
		3 & 0.45 & 1.15 $\pm$ 0.07 \\
		4 & 0.63 & 1.17 $\pm$ 0.08 \\
		5 & 0.81 & 1.24 $\pm$ 0.09 \\
		6 & 0.99 & 1.45 $\pm$ 0.09 \\
		7 & 1.17 & 1.47 $\pm$ 0.11 \\
		8 & 1.35 & 1.48 $\pm$ 0.16 \\
		9 & 1.53 & 1.84 $\pm$ 0.19 \\
		\noalign{\smallskip}\hline
		\hline
	\end{tabular}
\end{table}

In 2006, the Belle collaboration reported 
the measurements of $f_+^K(q^2)$ 
for $D^{0}\rightarrow K^{-}\ell^{+}\nu_{\ell}$ decays \cite{Belle2006}. Based on the accumulated $56461\pm309\pm830$ inclusive $D^0$ mesons, they found $1318\pm37\pm7$ signal events for the electron mode and $1249\pm37\pm25$ signal events for the muon mode. In neglecting the lepton masses, they obtained $f_+^K(q^2)$ in 27 $q^2$ bins 
with the bin size of 0.067 GeV$^2$. 
It is worthy to note that these measurements 
were obtained in the case of the masses of ignoring leptons, 
so the vector form factor is different from 
the one defined in this article.  
To make a distinction between these vector form factors, 
we use $f_+^{NL}(q^2)$ to represent the 
vector form factor in the case of neglecting
the mass of lepton. 
In order to use these measurements in this work, 
we translate them into 
products $f_+^{NL}(q_i^2)\rvert V_{cs} \rvert$ by 
using $\rvert V_{cs} \rvert=0.97296\pm0.00024$, 
which was used by the Belle experiment 
to obtain the $f_+^{NL}(q_i^2)$ in their article. 
The measurements $f_+^{NL}(q_i^2)$ and $f_+^{NL}(q_i^2)\rvert V_{cs} \rvert$ are listed in Tab. \ref{tabBelle}.

\begin{table}
	\caption{$f_+^{NL}(q_i^2)$ and $f_+^{NL}(q_i^2)\rvert V_{cs} \rvert$ measured at the Belle experiment.}
	\centering
	\begin{tabular}{rccc}
		\hline
		\hline\noalign{\smallskip}
		$i $        & $q^2$ (GeV) & $f_+^{NL}(q_i^2)$ & $f_+^{NL}(q_i^2)\rvert V_{cs} \rvert$ \\
		\noalign{\smallskip}\hline\noalign{\smallskip}
		1 & 0.100 & 0.711 $\pm$ 0.034 & 0.692 $\pm$ 0.033 \\
		2 & 0.167 & 0.787 $\pm$ 0.034 & 0.766 $\pm$ 0.033 \\
		3 & 0.233 & 0.764 $\pm$ 0.034 & 0.743 $\pm$ 0.033 \\
		4 & 0.300 & 0.838 $\pm$ 0.034 & 0.815 $\pm$ 0.033 \\
		5 & 0.367 & 0.788 $\pm$ 0.037 & 0.767 $\pm$ 0.036 \\
		6 & 0.433 & 0.843 $\pm$ 0.039 & 0.820 $\pm$ 0.038 \\
		7 & 0.500 & 0.882 $\pm$ 0.043 & 0.858 $\pm$ 0.042 \\
		8 & 0.567 & 0.942 $\pm$ 0.045 & 0.917 $\pm$ 0.044 \\
		9 & 0.633 & 0.910 $\pm$ 0.045 & 0.885 $\pm$ 0.044 \\
		10 & 0.700 & 0.823 $\pm$ 0.045 & 0.801 $\pm$ 0.044 \\
		11 & 0.767 & 1.028 $\pm$ 0.048 & 1.000 $\pm$ 0.047 \\
		12 & 0.833 & 1.000 $\pm$ 0.049 & 0.973 $\pm$ 0.048 \\
		13 & 0.900 & 0.949 $\pm$ 0.050 & 0.923 $\pm$ 0.049 \\
		14 & 0.967 & 1.046 $\pm$ 0.057 & 1.018 $\pm$ 0.055 \\
		15 & 1.033 & 1.100 $\pm$ 0.057 & 1.070 $\pm$ 0.055 \\
		16 & 1.100 & 0.941 $\pm$ 0.062 & 0.916 $\pm$ 0.060 \\
		17 & 1.167 & 1.114 $\pm$ 0.069 & 1.084 $\pm$ 0.067 \\
		18 & 1.233 & 1.100 $\pm$ 0.075 & 1.070 $\pm$ 0.073 \\
		19 & 1.300 & 1.249 $\pm$ 0.086 & 1.215 $\pm$ 0.084 \\
		20 & 1.367 & 1.381 $\pm$ 0.093 & 1.344 $\pm$ 0.090 \\
		21 & 1.433 & 1.313 $\pm$ 0.107 & 1.278 $\pm$ 0.104 \\
		22 & 1.500 & 1.190 $\pm$ 0.112 & 1.158 $\pm$ 0.109 \\
		23 & 1.567 & 1.416 $\pm$ 0.127 & 1.378 $\pm$ 0.124 \\
		24 & 1.633 & 1.471 $\pm$ 0.175 & 1.431 $\pm$ 0.170 \\
		25 & 1.700 & 1.417 $\pm$ 0.222 & 1.379 $\pm$ 0.216 \\
		26 & 1.767 & 1.150 $\pm$ 0.345 & 1.120 $\pm$ 0.336 \\
		27 & 1.833 & 1.450 $\pm$ 0.915 & 1.411 $\pm$ 0.890 \\
		\noalign{\smallskip}\hline
		\hline
	\end{tabular}
	\label{tabBelle}
\end{table}

$\mathfrak{f}_+^K(q^2)$ measured 
at the FOCUS experiment and $f_+^{NL}(q^2)|V_{cs}|$ measured 
at the Belle experiment are important for the determination of 
$f_0^{K}(q^2)$ for semileptonic $D \rightarrow K$ decays. 
We will discuss this issue in the next section.

\section{Fits to experimental data in the context of the SM}
\label{sec:4}
Our goal is to obtain the product $f_+^K(0)\rvert V_{cs}\rvert$ 
and the shapes of the vector and scalar form factors 
with semileptonic $D \to K$ decays from the existing experimental measurements. 
Firstly, we validate the analysis scheme by analyzing these 
experimental data, which is depending on the relative 
errors of these measurements and the contribution 
ratios of $f_0^K$ term to these measurements. 
If the contributions of $f_0^K$ to these measurements 
are much smaller than the 
errors of these measurements, the fitting result 
of the scalar form factor will not be credible, so the 
confirmation of the feasibility is very important. 

\subsection{The contribution of scalar form factor}
\label{sec:4.1}

\begin{figure*}[!]
	\centering
	\vspace*{8pt}
	\subfigure[]{
		\label{fig:subfig:a} 
		\includegraphics[width=2.8in]{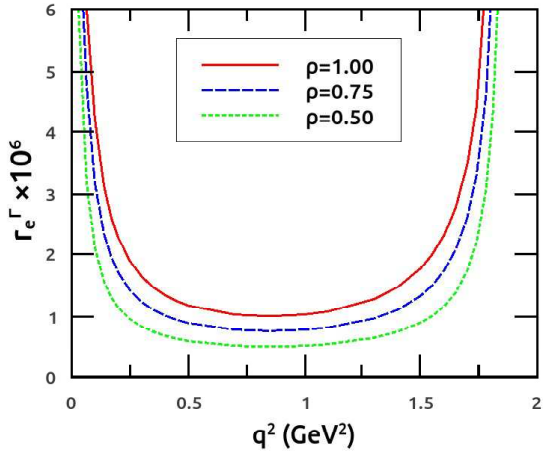}}
	\hspace{0.5in}
	\subfigure[]{
		\label{fig:subfig:b} 
		\includegraphics[width=2.8in]{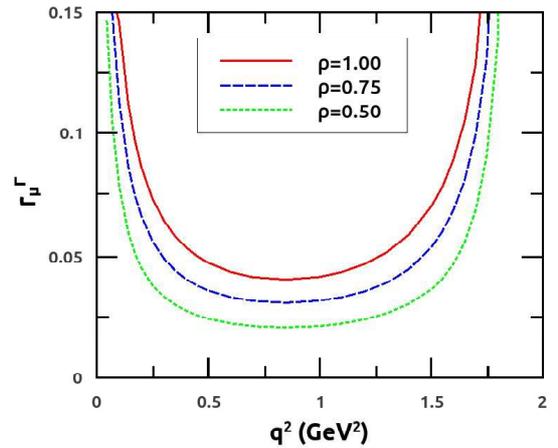}}
	\caption{The contributions of $f_0$ term to the differential decay rates for $D\rightarrow Ke^{+}\nu_{e}$ decays (a) and $D\rightarrow K\mu^{+}\nu_{\mu}$ decay (b).}
	\label{fig:efreu} 
\end{figure*}

The contribution of $f_0$ term to 
the differential decay rate of $D\rightarrow K\ell^{+}\nu_{\ell}$ at a certain $q^2$ can be described by the ratio
\begin{eqnarray}
r_{\ell}^{\Gamma}(q^2)&=& \dfrac{\mathrm{d}\Gamma_{\ell}^S/\mathrm{d}q^2}{\mathrm{d}\Gamma(D \rightarrow K \ell \nu_{\ell})/\mathrm{d}q^2} \nonumber \\
&=& \dfrac{\rho \dfrac{3m_{\ell}^{2}(m_{D}^{2}-m_{K}^{2})^{2}}{8 m_{D}^{2}\rvert \mathrm{\textbf{p}} \rvert^{2}q^{2}}}{1+\dfrac{m_{\ell}^{2}}{2q^{2}}+ \rho \dfrac{3m_{\ell}^{2}(m_{D}^{2}-m_{K}^{2})^{2}}{8 m_{D}^{2}\rvert \mathrm{\textbf{p}} \rvert^{2}q^{2}}},
\label{eq:efreu}
\end{eqnarray}
where $\rho$ represents $\rvert f_0^K(q^2)/f_+^K(q^2) \rvert^2$.

$r_{e}^{\Gamma}(q^2)$ and $r_{\mu}^{\Gamma}(q^2)$ varying with $q^2$ were shown in Figs. \ref{fig:efreu} (a) and (b), respectively. 
The $\rho$ in Eq. \ref{eq:efreu} is set to $\rho=$ 1.0, 0.75 or 0.5 according to previous lattice calculations. 
Fig. \ref{fig:efreu} (a) shows that the contribution of $f_0^K$ term to the partial 
decay rate for $D\rightarrow K e^+ \nu_{e}$ decay is less than $10^{-5}$ in the most range of $q^2$,
which is much smaller than the relative errors of corresponding partial decay rate measured at the experiments 
listed in Tabs. \ref{tabD0} and \ref{tabD1}, so neglecting the contribution of the scalar form factor is a good approximation in analysis
for the electron channel.
While for the muon channel, Fig. \ref{fig:efreu} (b) 
shows that the contribution of $f_0^K$ term to the partial decay rate 
is 3\%$\sim$5\% in the most $q^2$ range which need to be considered when the experimental measurements have high precision.
So the extraction of $f_0^K(q^2)$ is feasible especially from the muon channel.

\subsection{Construct Chi-squared function}
\label{sec:4.2}
To obtain $f_+^K(0)\rvert V_{cs}\rvert$ and shapes of the vector and scalar form factors, we perform our fit to these experimental measurements by minimizing the Chi-squared function

\begin{eqnarray}
\chi^2=\chi_{\Delta\Gamma}^2+\chi_{FOC}^2 +\chi_{Bel}^2,
\label{chitotal}
\end{eqnarray}
where $\chi_{\Delta\Gamma}^2$ is constructed for measurements of partial decay rates in different $q^2$ ranges for $D\rightarrow K e^+\nu_{e}$ as shown in Tabs. 3 and 4, $\chi_{FOC}^2$ is for the non-parametric form factors $\mathfrak{f}_+^K(q^2)$ measured at the FOCUS experiment, and $\chi_{Bel}^2$ corresponds to the Belle Collaboration measured products $f_+^{NL}(q_i^2)\rvert V_{cs} \rvert$.

Since there are correlations between the measurements of partial decay rates for $D^{0}\rightarrow K^{-}e^{+}\nu_{e}$ decays and/or $D^{+}\rightarrow \bar{K}^{0}e^{+}\nu_{e}$ decays, the $\chi_{\Delta\Gamma}^2$ is given by
\begin{equation}
\chi_{\Delta\Gamma}^2 = \sum_{i=1}^{62} \sum_{j=1}^{62} (\Delta\Gamma_i-\Delta\Gamma_i^{th})(C_{\Delta\Gamma}^{-1})_{ij}(\Delta\Gamma_j-\Delta\Gamma_j^{th}),
\end{equation}
where $\Delta\Gamma$ is the partial decay rate measured in experiment, $\Delta\Gamma^{th}$ denotes its theoretical expectation, and $C_{\Delta\Gamma}^{-1}$ is the inverse of the covariance matrix $C_{\Delta\Gamma}$, which is a $62\times62$ matrix. To compute the covariances of these 62 partial decay rates measured in different $q^2$ ranges and at different experiments, we adopt the concept proposed in Ref. \cite{Fang2014}: (a) at the same experiment, the statistical and systematic errors of these partial decay rates, and corresponding correlations between these partial decay rates are used to compute their covariances; (b) the systematic uncertainties caused by the lifetime of $D^{0(+)}$ meson are fully correlated among all of the partial decay rates for $D^{0(+)} \rightarrow K^-(\bar{K}^0) e^+ \nu_{e}$ decays measured at different experiments. (c) the systematic uncertainties related to $D^{0}\rightarrow K^{-}\pi^{+}$ are full correlated among all of the measurements of category (i) in Section 3 for $D^{0} \rightarrow K^- e^+ \nu_{e}$ decays. 

Due to the correlations between measurements of the non-parametric form factors at the FOCUS experiment, the $\chi_{FOC}^2$ in Eq. \ref{chitotal} is defined as
\begin{eqnarray}
\chi_{FOC}^2 = \sum_{i=1}^{9} \sum_{j=1}^{9} (\mathfrak{f}_i-\mathfrak{f}_i^{th})(C_{FOC}^{-1})_{ij}(\mathfrak{f}_j-\mathfrak{f}_j^{th}),
\label{eq:chiFOC}
\end{eqnarray}
where $\mathfrak{f}_i$ and $\mathfrak{f}_i^{th}$ are the measured value from the FOCUS experiment and the theoretical expectation of the average of $\mathfrak{f}_+^K(q^2)$ over the width of $i$-th $q^2$ bin, respectively. It is worth noting that vector form factor $f_+^K(q^2)$ in Eq. \ref{form factor 2} can not be used as the theoretical form factor $\mathfrak{f}_i^{th}$ directly, because some assumptions about the expression of differential decay rate for $D^{0}\rightarrow K^{-}\mu^{+}\nu_{\mu}$ process are different between this article and Ref. \cite{FOCUS-2004}. By comparing Eq. (\ref{eq:DKSM}) in this article and Eq. (2) in Ref. \cite{FOCUS-2004}, we can obtain
\begin{equation}
\mathfrak{f}_i^{th} = \frac{ \left[\int_{q_{i\,min}^2}^{q_{i\,max}^2}\dfrac{ V_\mu(q^{2}) \rvert f_{+}^K(q^{2}) \rvert^{2} 
		+ S_\mu(q^{2}) \rvert f_{0}^K(q^{2}) \rvert^{2}}
	{ V_\mu(q^{2}) + S_\mu(q^{2}) \left( 1+\beta q^2/\alpha \right)^2 }\mathrm{d}q^2 \right]^{\frac{1}{2}}}{f_+^K(0)\sqrt{q_{i\,max}^2-q_{i\,min}^2}},
\label{eq:ffoc}
\end{equation}
where $S_{\mu}(q^2)=3m_{\mu}^{2}(m_{D}^{2}-m_{K}^{2})^{2}/(8 m_{D}^{2}\rvert \mathrm{\textbf{p}} \rvert^{2}q^{2})$, $V_{\mu}(q^2)=\left(1+m_{\mu}^{2}/(2q^{2}) \right)$, $\alpha=(m_{D}^{2}-m_{K}^{2})^2$ and $\beta=f_-^K(q^2)/f_+^K(q^2)\\=-0.7$. The $C_{FOC}^{-1}$ in Eq. (\ref{eq:chiFOC}) is the inverse of the covariance matrix $C_{FOC}$, which is a $9\times9$ matrix. We can construct the covariance matrix $C_{FOC}$ by the relation $(C_{FOC})_{ij}=\sigma_i\sigma_j\rho_{ij}$, where $\sigma_{i}$ ($\sigma_{j}$) is the standard error of $\mathfrak{f}_+^K(q^2)$ at the central value of the $i$-th ($j$-th) $q^2$ bin measured at the FOCUS experiment, and $\rho_{ij}$ is the correlation coefficient of measurements of $\mathfrak{f}_+^K(q^2)$ at $i$-th $q^2$ bin and $j$-th $q^2$ bin.

The $\chi_{Bel}^2$ in Eq. (\ref{chitotal}) is built for the products $f_+^{NL}(q_i^2)\\\rvert V_{cs} \rvert$ measured at the Belle experiment. The $\chi_{Bel}^2$ is defined as
\begin{eqnarray}
\chi_{Bel}^2 = \sum_{i=1}^{27}\left(\frac{F_i-F_i^{th}}{\sigma_i} \right)^2,
\label{eq:chiBel}
\end{eqnarray}
where $F_i$ and $F_i^{th}$ are experimental and theoretical values of $f_+^{NL}(q_i^2)\rvert V_{cs} \rvert$ in the $i$-th $q^2$ bin respectively, and $\sigma_i$ represents the standard deviation of $F_i$. In Eq. \ref{eq:chiBel}, we neglect some possible correlations among the measurements of $f_+^{NL}(q_i^2)\rvert V_{cs} \rvert$. Similar to the analysis of the measurements at the FOCUS experiment above, by comparing Eq. (\ref{eq:DKSM}) in this article and Eq. (1) in Ref. \cite{Belle2006}, the expression of $f_+^{NL}(q_i^2)\rvert V_{cs} \rvert$ is

\begin{eqnarray}
F_i^{th} = \left[ \dfrac{\int_{q_{i\,min}^2}^{q_{i\,max}^2} \left(0.54\dfrac{ \mathrm{d}\Gamma_e}{\mathrm{d}q^2}  + 0.46 \dfrac{ \mathrm{d}\Gamma_\mu}{\mathrm{d}q^2} \right) \mathrm{d}q^2}{q_{i\ max}^2-q_{i\ min}^2} \dfrac{24\pi^3}{G_F^2|\mathrm{\textbf{p}_i}|^3} \right]^{1/2}, \nonumber \\
\label{Fith}
\end{eqnarray}
where $ \mathrm{d}\Gamma_e/\mathrm{d}q^2$ and $ \mathrm{d}\Gamma_\mu/\mathrm{d}q^2$ are respectively the Eq. (\ref{eq:DKSM}) for $D\rightarrow K e^+ \nu_{e}$ and $D\rightarrow K \mu^+ \nu_{\mu}$ decays. The weights 0.54 and 0.46 are obtained from the branching fractions of $D^0 \rightarrow K^- e^+ \nu_{e}$ and $D^0 \rightarrow K^- \mu^+ \nu_{\mu}$ and their errors which are in the Belle’s paper published. In Eq.(\ref{Fith}),
\begin{eqnarray}
|\mathrm{\textbf{p}_i}|^3=\frac{\int_{q_{i\,min}^2}^{q_{i\,max}^2}|\mathrm{\textbf{p}}|^3 |f_+^K(q^2)|^2 \mathrm{d}q^2}{|f_+^K(q_i^2)|^2(q_{i\ max}^2-q_{i\ min}^2)},
\end{eqnarray}
where $f_+^K(q^2)$ is computed using the simple pole model with $m_{pole}=1.82\pm0.04\pm0.03$ GeV which was originally used to obtain $f_+^K(q^2)$ in the Belle's paper published.

\subsection{Fit to experimental data}

\begin{table*}
	\caption{The extracted values of the parameters $f_+^K(0)|V_{cs}|$, $r_{+1}$, $r_{+2}$ and $r_{01}$ and the correlation coefficients betwee them $\rho_{ij}$ in the fittings where i and j with the values from 1 to 4 correspond to the four parameters. $r_{+1}$ and $r_{+2}$ are the shape parameters of $f_+(q^2)$, and $r_{01}$ is of $f_0(q^2)$. The $N_V$ and $N_S$ are the expansion orders of the vector and scalar form factors, respectively. The goodness of fit $\chi^2$/$\mathrm{d}.\mathrm{o}.\mathrm{f}.$ is also listed. "-" means unavailable. }
	\label{fitresult}
	\centering
	\begin{tabular}{c|c|c|c|c|c}
		\hline
		\hline\noalign{\smallskip}
		\multicolumn{1}{c|}{\multirow{4}{*}} & \multicolumn{3}{c|}{\multirow{2}{*}{This work}} & HFLAV'16 & Y.~Fang  \\
		\multicolumn{1}{c|}{}& \multicolumn{1}{c}{} &\multicolumn{1}{c}{} & \multicolumn{1}{c|}{} & & \\
		\cline{2-6}
		&(a) $N_V=1$\ $N_S=1$ &(b) $N_V=2$ &(c) $N_V=2$\ $N_S=1$ & $N_V=2$ & $N_V=2$ \\
		& $m_{\ell}\neq0$ & $m_{\ell}=0$ & $m_{\ell}\neq0$ & $m_{\ell}=0$ & $m_{\ell}=0$ \\
		\hline
		$f_+^K(0)|V_{cs}|$ & 0.7182(29) & 0.7169(29) & 0.7182(34) & 0.7226(32) & 0.717(4) \\
		& & & & \\
		$r_{+1}$ & -2.16(7) & -2.13(11) & -2.16(11) & -2.38(13) & -2.34(17) \\
		& & & & \\
		$r_{+2}$ & $-$ & -0.84(2.20) & -0.07(2.72) & -4.7(3.0) & 0.43(3.82) \\
		& & & & \\
		$r_{01}$ & 0.89(3.27) & $-$ & 0.84(3.73) & $-$ & $-$  \\
		\hline
		\multirow{4}{*}{Correlations}& & & $\rho_{12}/\rho_{13}/\rho_{14}/$ & &\\
		& $\rho_{12}/\rho_{14}/\rho_{24}$ & $\rho_{12}/\rho_{13}/\rho_{23}$  & $\rho_{23}/\rho_{24}/\rho_{34}$  & $\rho_{12}/\rho_{13}/\rho_{23}$ & \\
		& 0.52/-0.32/-0.71 & -0.007/0.22/-0.87 & -0.02/0.39/-0.39/ & -0.19/0.51/-0.84 &$-$ \\
		& & & -0.79/0.10/-0.37 & \\
		\hline
		$\chi^2$/$\mathrm{d}.\mathrm{o}.\mathrm{f}.$ & 92.334/95 & 99.2/95 & 92.331/94 & $-$ & 100.1/69 \\
		\noalign{\smallskip}\hline
		\hline
	\end{tabular}
\end{table*}

We fit the experimental data with the Eq.1 where the vector and 
scalar form factors are parameterized as Eq. (\ref{form factor 2}). 
The contribution of $f_0^K(q^2)$, as shown in Sec.\ref{sec:4.1}, is 
relatively small ($3\%\sim5\%$) for the 
muon case and negligible ($\sim10^{-4}\%$) 
for electron due to their small mass, so the extraction of $f_0^K(q^2)$ is 
sensitive to the parameterization of $f_+^K(q^2)$. 
The fittings by expanding $f_+^K(q^2)$ 
to different orders and negelecting the mass of lepton or not are performed.  
The three fitting schemes are applied,  
(a) $N_V=1$, $N_S=1$ 
and $m_{\ell}\neq0$; (b) $N_V=2$ and $m_{\ell}=0$; 
(c) $N_V=2$, $N_S=1$ and $m_{\ell}\neq0$. 
$N_V$ and $N_S$ are the expansion orders 
of $f_+^K(q^2)$ and $f_0^K(q^2)$ in Eq. (\ref{form factor 2}), 
respectively. All of the SM input parameters 
such as the Fermi constant $G_F$, 
the masses of mesons and charged leptons 
are taken from PDG \cite{PDG}. 

The fitting results are listed in the Tab. \ref{fitresult}. 
As a comparison, the results obtained by Heavy Flavor Averaging Group 
(HFLAV) \cite{HFLAV2016} in 2016 (HFLAV'16) and Y. Fang et al \cite{Fang2014}
are also listed.
The experimental data applied by HFLAV'16, Y. Fang
and our work are a bit different.
Comparing to the work of HFLAV'16 and Y.~Fang, 
the latest results of 
the $D^+ \rightarrow \bar{K}^0 e^+ \nu_{e}$ from 
BESIII\cite{BES-III-2016,BES-III-2017} is included in our analysis. 
In addition, in order to extract $f_0^K(q^2)$ effectively, 
more measurements for muon channel are added in our work e.g. the  
total decay rates of $D \rightarrow K \mu^+ \nu_{\mu}$
measured by E686\cite{E6871993,E6871995}, 
E653\cite{E653}, CLEO\cite{CLEO-1991}, FOCUS\cite{FOCUS20041}, 
BES-II\cite{BESII2006,BESII20061}, BES-III\cite{BES-III-2016mu}.
The measurement from CLEO-c Ref.\cite{CLEO-c-2008} is used in HFLAV'16
only.

\begin{figure*}[!]
	\centering
	\vspace*{8pt}
	\subfigure[]{
		\label{fig:subfig:a} 
		\includegraphics[width=2.0in]{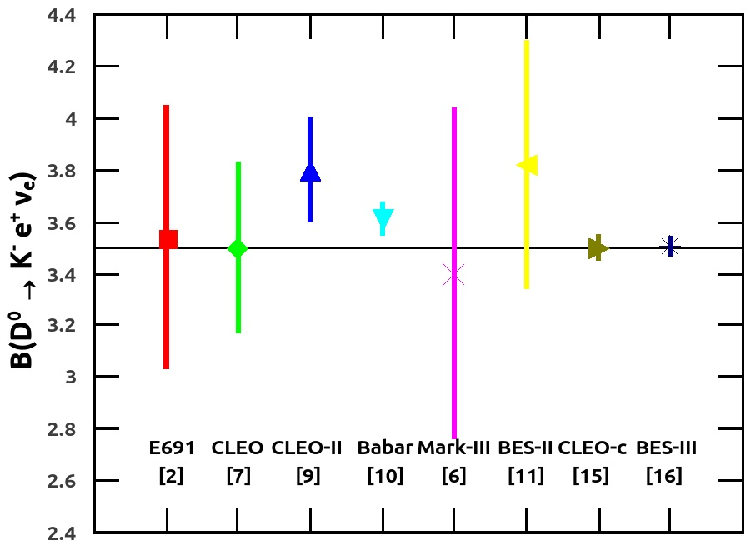}}
	\hspace{0.1in}
	\subfigure[]{
		\label{fig:subfig:b} 
		\includegraphics[width=2.0in]{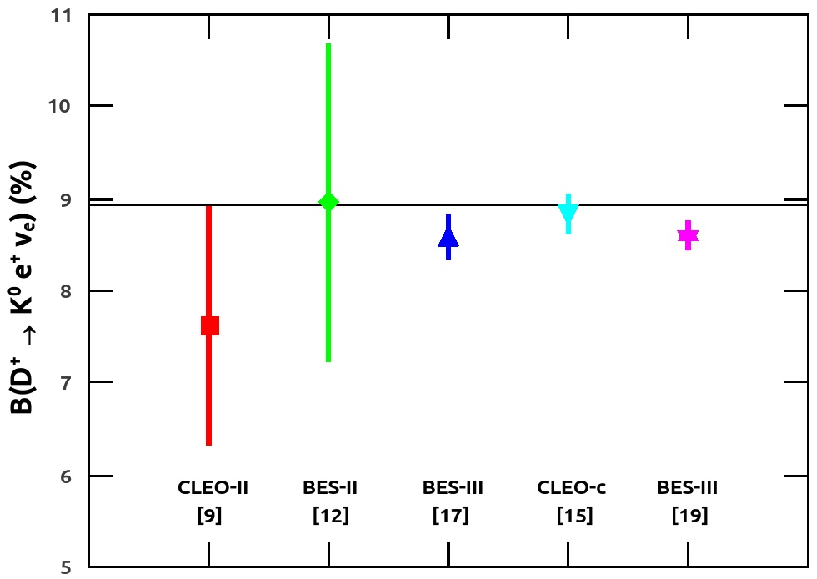}}
	\hspace{0.1in}
	\subfigure[]{
		\label{fig:subfig:c} 
		\includegraphics[width=2.0in]{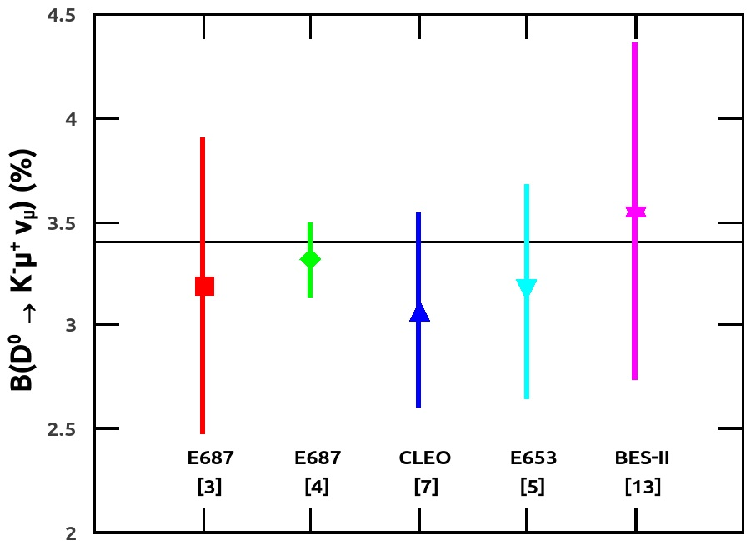}}
	\hspace{0.1in}
	\subfigure[]{
		\label{fig:subfig:d} 
		\includegraphics[width=2.0in]{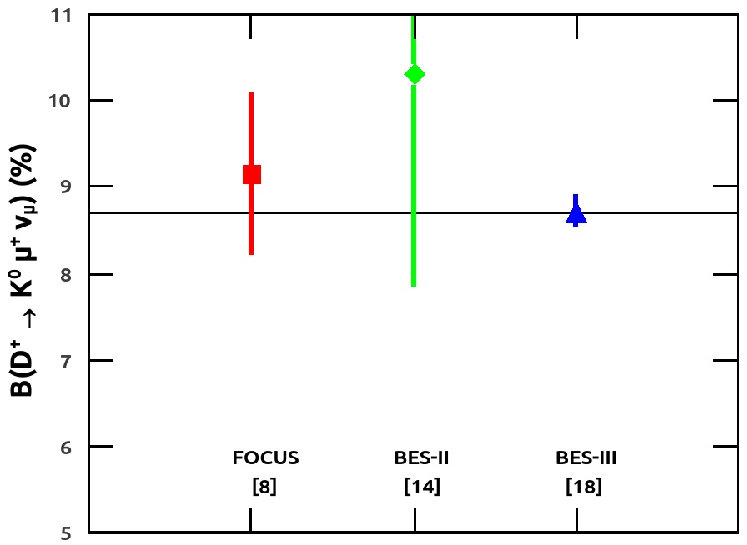}}
	\hspace{0.1in}
	\subfigure[]{
		\label{fig:subfig:e} 
		\includegraphics[width=2.0in]{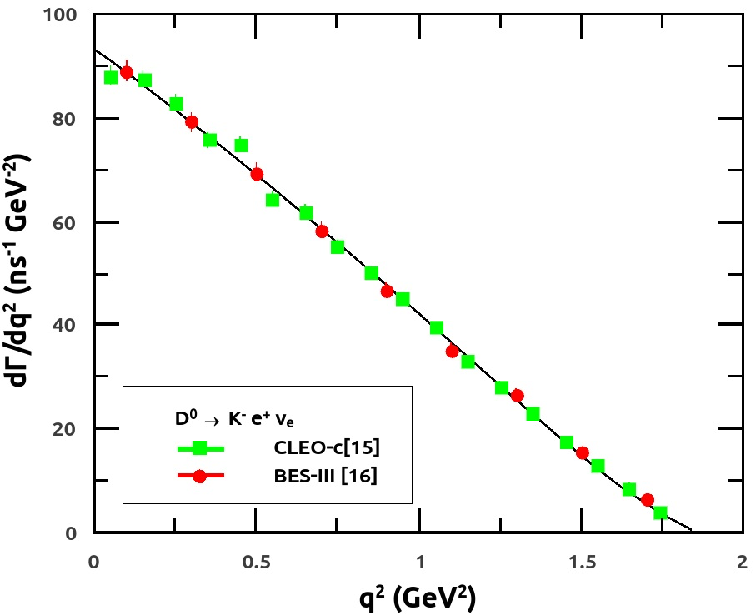}}
	\hspace{0.1in}
	\subfigure[]{
		\label{fig:subfig:f} 
		\includegraphics[width=2.0in]{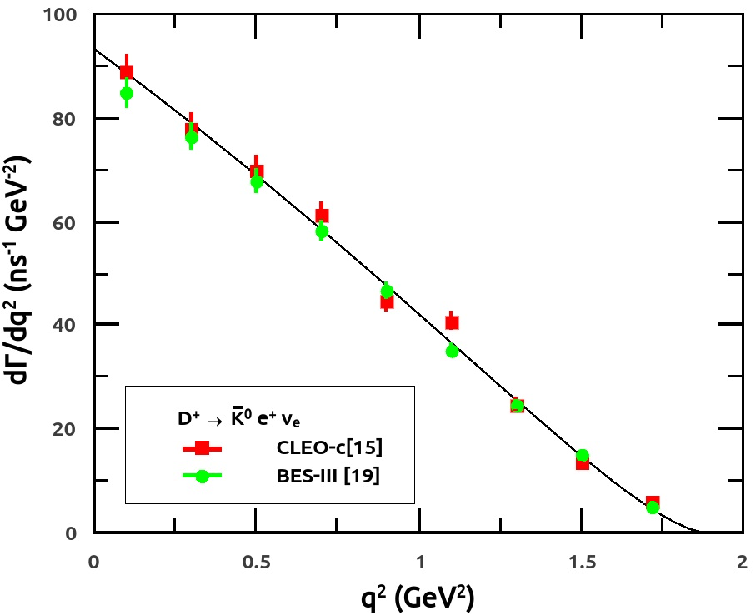}}
	\hspace{0.1in}
	\subfigure[]{
		\label{fig:subfig:g} 
		\includegraphics[width=2.0in]{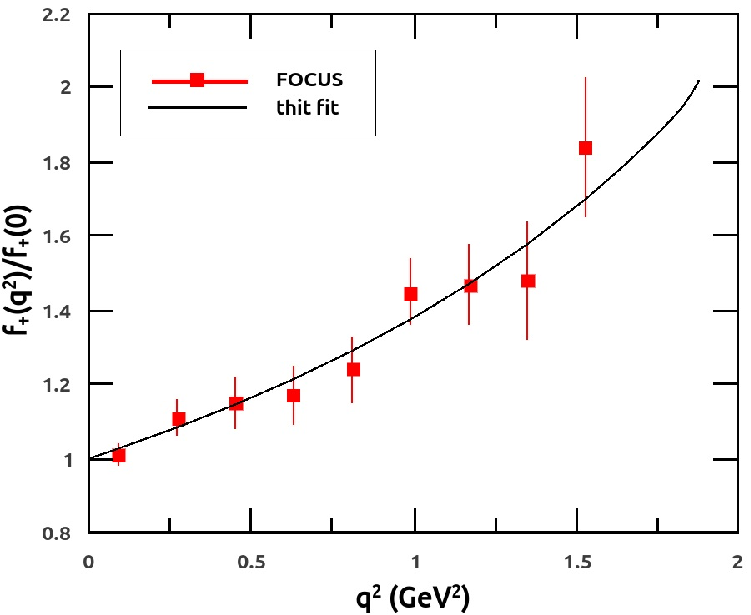}}
	\hspace{0.1in}
	\subfigure[]{
		\label{fig:subfig:h} 
		\includegraphics[width=2.0in]{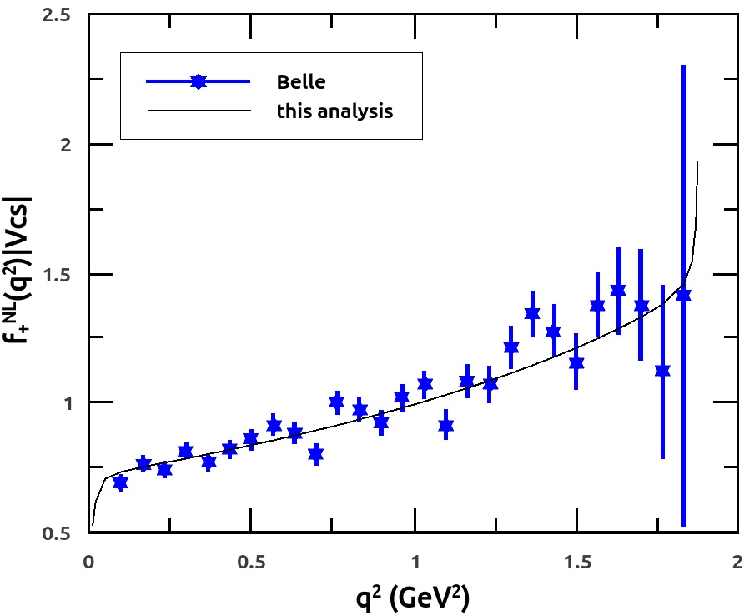}}
	\hspace{0.1in}
	\subfigure[]{
		\label{fig:subfig:i} 
		\includegraphics[width=2.0in]{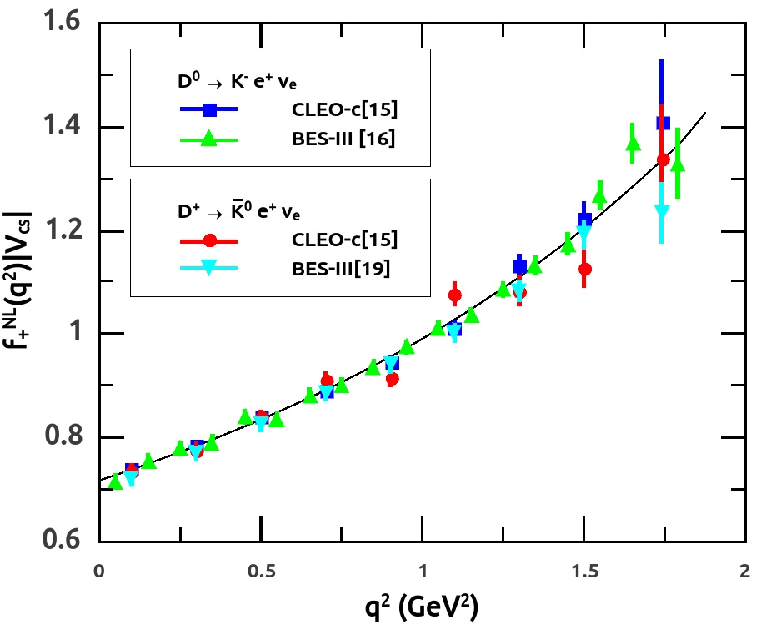}}
	\caption{Fit to the experimental measurements. The black lines are the fitting results, and the different symbols are the measurements from different experiments. The branching fractions for $D^0\rightarrow K^- e^+ \nu_{e}$ (a), $D^+ \rightarrow \bar{K}^0 e^+\nu_{e}$ (b), $D^0\rightarrow K^- \mu^+ \nu_{\mu}$ (c) and $D^+ \rightarrow \bar{K}^0 \mu^+\nu_{\mu}$ (d) at different experiments. Differential decay rates measured at the CLEO-c and BES-III experiments for $D^0\rightarrow K^- e^+ \nu_{e}$ (e) and $D^+ \rightarrow \bar{K}^0 e^+\nu_{e}$ (f). The relative form factor $\mathfrak{f}_+^K(q^2)$ measured at the FOCUS experiment for $D^0\rightarrow K^- \mu^+ \nu_{\mu}$ decay (g). The product $f_+^{NL}(q^2)|V_{cs}|$ measured at the Belle experiment for $D^0\rightarrow K^- \ell^+ \nu_{\ell} (l=e,\mu)$ decay (h). The product $f_+^{NL}(q^2)|V_{cs}|$ measured at the CLEO-c and the BES-III experiments for $D\rightarrow K e^+ \nu_{e}$ decay (i). }
	\label{fig:fitSM} 
\end{figure*}

The parameter setting in the fittings of HFLAV'16 and Y.~Fang is to expand 
$f_+^K(q^2)$ with two orders and neglect the lepton mass, which is the
same as the fitting scheme (b) in our work.
From the Tab.~\ref{fitresult} we can see that the $f_+^K(0)|V_{cs}|$
obtained by HFLAV'16, Fang and our work (b) are consistent within error.
The values of the shape parameter $r_{+1}$ and $r_{+2}$ are also consistent 
within two times of the errors, while $r_{+2}$ has quite large errors. 
Comparing the results of the schemes (a), (b) and (c) of our work in Tab.~\ref{fitresult}
we can see
that the fitting results have little change with the different settings of the  
$f_+^K(q^2)$ expansion and the lepton mass. While the $\chi^2$/$\mathrm{d}.\mathrm{o}.\mathrm{f}.$ become slightly
better when the lepton mass is not neglected.
The fit of the scheme (a) is taken as the nominal one, and 
the corresponding fitting results are shown in the Fig. \ref{fig:fitSM}.

\subsection{Determinations of $f_+^K(0)$ and $\rvert V_{cs} \rvert$}

\begin{figure}[!]
	\centerline{\includegraphics[width=3.0in]{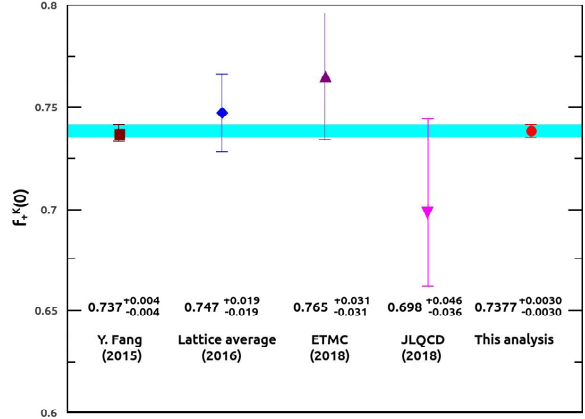}}
	\vspace*{8pt}
	\caption{The $f_+^K(0)$ determined with experimental data or Lattice QCD. The values are described in the text. The cyan band is the uncertainty at one standard deviation of this analysis. \protect\label{fig:comparef0}}
\end{figure}

Note that the fit to experimental data returns just the product of the hadronic form factor $f_+^K(0)$ and the magnitude of CKM matrix element $V_{cs}$. To determine $f_+^K(0)$ or extract $\rvert V_{cs} \rvert$, we need more inputs. $\rvert V_{cs} \rvert$ can be most precisely determined using a global fit to all available measurements and take three generation unitarity as the SM constrain. The theory predictions for hadronic matrix elements is also needed in the fit. There are several approaches to combining the experimental data. By considering the product $f_+^K(0)\rvert V_{cs} \rvert=0.7182 \pm 0.0029$ as shown in Tab. \ref{fitresult} together with $\rvert V_{cs} \rvert=0.97351\pm0.00013$ obtained from the unitarity constraints\cite{PDG}, one can obtain

\begin{eqnarray}
f_+^K(0)=0.7377 \pm 0.0030 \pm 0.000,
\end{eqnarray}
where the first error is from the uncertainties in the partial decay rate measurements, and the second is the contribution of the uncertainty of $\rvert V_{cs} \rvert$. The form factor $f_+^K(0)$ determined from recent lattice calculations by the JLQCD collaboration \cite{JLQCD} and the ETMC collaboration \cite{ETM}, the average of $N_f=2+1$ lattice calculations before 2017 \cite{LA2016} and experimental fit in 2014 \cite{Fang2014} are compared in Fig. \ref{fig:comparef0}. Our fitting result is consistent with these theoretical calculations and presents a good consistency with the previous fitting result, but is with higher precision. $f_+^K(0)$ also can be determined from such as light-cone sum rules\cite{Khodjamirian:2009ys} and the light front quark model\cite{Cheng:2017pcq}.

\begin{figure*}[!]
	\centering
	\vspace*{0cm}
	\subfigure[]{
		\label{fig:subfig:a} 
		\includegraphics[width=2.8in]{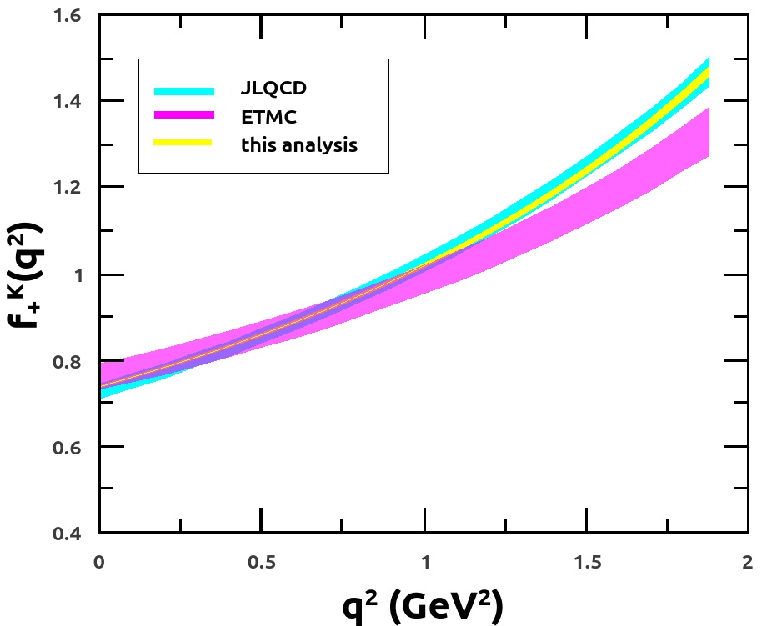}}
	\hspace{0.5in}
	\subfigure[]{
		\label{fig:subfig:b} 
		\includegraphics[width=2.8in]{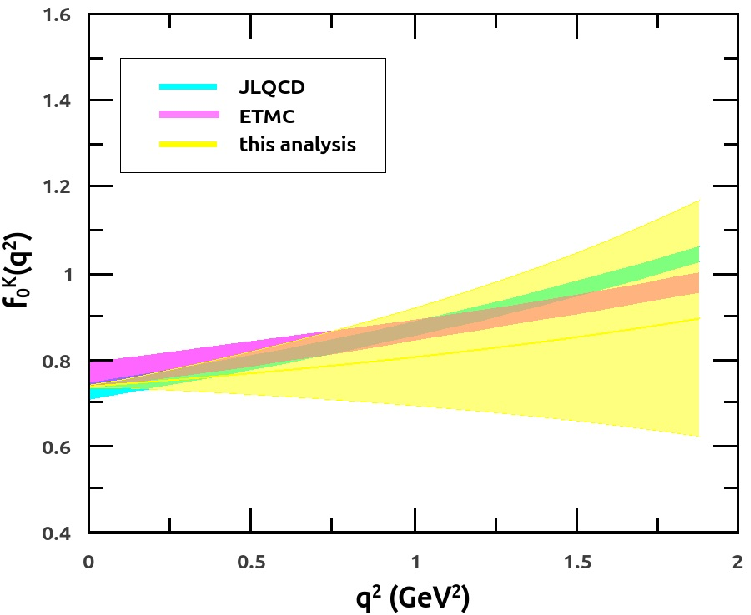}}
	\caption{The shapes of $f_+^K(q^2)$ (a) and $f_0^K(q^2)$ (b). The yellow bands are the results of this work. The cyan and pink bands are from the JLQCD\cite{JLQCD} and ETMC\cite{ETM}.}
	\label{fig:LQCDff} 
\end{figure*}

The $f_+^K(q^2)$ and $f_0^K(q^2)$ for   
semileptonic $D\rightarrow K$ decays are determined
and shown in Fig. \ref{fig:LQCDff}.
The recent lattice calculation from JLQCD collaboration 
with $N_f=2+1$ flavors of dynamical quarks \cite{JLQCD}
and ETMC collaboration \cite{ETM} with $N_f=2+1+1$ are shown in the plot as well.
We can see that the $f_+^K(q^2)$ and $f_0^K(q^2)$ obtained by this work agree well with
the JLQCD result within error, and also 
the ETMC result at low values of $q^2$. 
There are some descripancies at high values of $q^2$
between the ETMC and the other two results which is caused by
the subtraction
of hypercubic artifacts in the ETMC calculation.
The hypercubic effects will impact the form factors especially at the high 
values of $q^2$ \cite{ETM}.
The precision of the $f_0^K(q^2)$ obtained in the work is low due to 
the small contribution of the scalar form factor to the decay width as discussed in Section \ref{sec:4.1}, which 
is expected to be improved according to more measurements
of the $D\rightarrow K\mu\nu_\mu$ decay.


\begin{figure*}[!]
	\centerline{\includegraphics[width=5.0in]{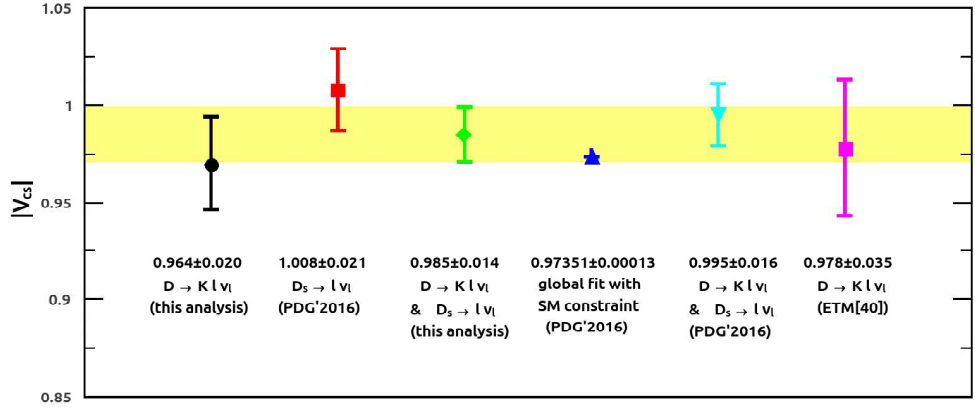}}
	\vspace*{8pt}
	\caption{The $\rvert V_{cs} \rvert$ extracted from different analysis. The yellow band is one standard deviation of the green square \protect\label{fig:Vcs}}
\end{figure*}

On the other hand, a comprehensive consideration of $f_+^K(0)\rvert V_{cs} \rvert=0.7182 \pm 0.0029$ and $f_+^K(0)=0.745\pm0.015$ (the average of lattice calculations $f_+^K(0)=0.698\pm0.041$ obtained by JLQCD collaboration\cite{JLQCD}, $f_+^K(0)=0.765\pm0.031$ obtained by ETMC \cite{ETM} and $f_+^K(0)=0.747\pm0.019$ which is the lattice average before 2017\cite{LA2016}), the magnitude of the CKM matrix element $V_{cs}$ is determined to be
\begin{eqnarray}
\rvert V_{cs} \rvert=0.964\pm0.004\pm0.019,
\end{eqnarray}
where the first error is experimental and the second is theoretical. With comprehensive consideration of the $\rvert V_{cs} \rvert=0.964\pm0.004\pm0.019$ extracted in this analysis together with the $\rvert V_{cs} \rvert=1.008\pm0.021$ determined from leptonic $D_s$ decay \cite{PDG}, the magnitude of the CKM matrix element $V_{cs}$ is determined to be
\begin{eqnarray}
\rvert V_{cs} \rvert=0.985\pm0.014,
\end{eqnarray}

which is consistent with the average value $\rvert V_{cs} \rvert=0.995\pm0.016$ from PDG'2016 \cite{PDG}. The ETM collaboration obtained $\rvert V_{cs} \rvert=0.978\pm0.035$ by combining $f_+^K(q^2)$ obtained by their lattice QCD simulations with the differential rates measured for the semileptonic $D\rightarrow K \ell \nu_{\ell}$ decays\cite{Riggio:2017zwh}. Comparisons of $\rvert V_{cs} \rvert$ extracted from different analysis are shown in Fig. \ref{fig:Vcs}.

\section{Leptoquark constraints from $D\rightarrow K\ell^+\nu_{\ell}$ decays}
\label{sec:5}
If there exist new interactions beyond the Standard $W^+$ mediating $c\bar{s} \rightarrow v\bar{\ell}$, then they alter the decay rate and $q^2$-distribution of $ D\rightarrow K \ell \nu_{\ell} $ decay processes. Since experimental data for $ D\rightarrow K \ell \nu_{\ell} $ decays, at some level, are in good agreement with the SM, the contributions of the new particles should be small and thus these new particles must be too heavy to directly detect. So an effective Lagrangian extended the SM is considered here. Omitting right-handed neutrinos, the effective Lagrangian is given by \cite{Kronfeld2008}
\begin{eqnarray}
\mathcal{L}_{eff}=&-&( \sqrt{2}G_F V_{cs}^*+G_V)(\bar{s} \gamma^\mu c)(\bar{\nu}_L \gamma_\mu l_L) \nonumber \\
&+& ( \sqrt{2}G_F V_{cs}^*+G_A)(\bar{s} \gamma^\mu \gamma_5 c)(\bar{\nu}_L \gamma_\mu l_L) \nonumber \\
&-& G_S(\bar{s} c)(\bar{\nu}_L l_R)-G_P(\bar{s}\gamma_5 c)(\bar{\nu}_L l_R) \nonumber \\
&-& G_T(\bar{s} \sigma^{\mu\nu} c)(\bar{\nu}_L \sigma_{\mu\nu} l_R)+\mathrm{h.c.},
\label{LNP1}
\end{eqnarray}  
where the terms parameterized by $G_F V_{cs}^*$ represent the effective SM Lagrangian, the other five arose by new type interactions. 

A few possibilities can arise the non-Standard contributions to charm meson leptonic and
semileptonic decays which have been analyzed in Refs. \cite{Dobrescu2008,Kronfeld2008,Barranco2016,Dorsner:2009cu,Fajfer:2015ixa}. Other attempts
to account for flavour symmetry breaking
in pseudoscalar meson decay constants previously presented in Refs. \cite{Gershtein:1976aq,Khlopov:1978id,Komachenko:1983jv,Komachenko:1987gg}. As an example among the candidates which can lead to an effective $\bar{s}c\bar{\nu}\ell$ vertex they discussed, the mechanism of the $u$-channel exchange of a charge $-1/3$ scalar leptoquark $S_0$ is analyzed here. $S_0$ transforms as color-triplet and weak-singlet with the $U(1)$ hyper-charge $-2/3$ under $SU(3)_c\times SU(2)_L\times U(1)_Y$ transformations of the SM\cite{Kronfeld2008,LQ-review}. The interactions between $S_0$ and the SM fermions can be described as\cite{Kronfeld2008}
\begin{eqnarray}
\mathcal{L}_{S_0}= \lambda_{2i}^{LL}(\bar{c}_Ll_{iL}^C-\bar{s}_L\nu_{iL}^C)S_0+\lambda_{2i}^{RR}\bar{c}_Rl_{iR}^CS_0+\mathrm{h.c.},
\label{LQ1/3}
\end{eqnarray}  
where the superscript $C$ stands for charge conjugation, the subscript $i$ denotes the generation of lepton, and $\lambda_{2i}^{LL}$ and $\lambda_{2i}^{RR}$ are complex Yukawa couplings. When the leptoquark mass satisfy $m_{S_0}\gg m_D$, one can obtain
\begin{eqnarray}
&&G_V=G_A=\frac{\rvert \lambda_{2i}^{LL} \rvert^2}{4m_{S_0}^2}, \label{lq21} \\
&&G_P=G_S=\frac{\lambda_{2i}^{LL} \lambda_{2i}^{RR*}}{4m_{S_0}^2}=-2G_T. \label{lq22}
\end{eqnarray}
Then, there are only two unrelated coefficients left for two type fermion-leptoquark-fermion interactions in Eq. (\ref{LNP1}). We can name the one parameterized by $\rvert \lambda_{2i}^{LL} \rvert^2/m_{S_0}^2$ the $LL$ type, which is caused by only the scalar leptoquark $S_0^L$, and the one parameterized by $\lambda_{2i}^{LL} \lambda_{2i}^{RR*}/m_{S_0}^2$ the $LR$ type.

Because of these new interactions, the expression of differential decay rate for $ D\rightarrow K \ell \nu_{\ell} $ decays, Eq. (\ref{eq:DKSM}), should be rewritten as
\begin{eqnarray}
\label{eq:DKNP}
\dfrac{\mathrm{d}\Gamma_{\ell}^{NP}}{\mathrm{d}q^{2}}&=& \frac{\rvert \mathrm{\textbf{p}} \rvert^3}{24\pi^{3}} \left(1-\frac{m_{\ell}^2}{q^2} \right)^2 \cdot \left\{ \left( 1+\dfrac{m_{\ell}^{2}}{2q^{2}} \right) \left|\left(G_FV_{cs}^*+\frac{G_V}{\sqrt{2}} \right) f_{+}^K(q^{2}) \right|^{2} \nonumber
\right. \\
&+& \left. \frac{3(m_{D}^{2}-m_{K}^{2})^{2}}{8m_{D}^{2}\rvert \mathrm{\textbf{p}} \rvert^{2}}\rvert f_{0}^K(q^{2}) \rvert^{2} \cdot \left|\frac{m_{\ell}}{\sqrt{q^{2}}}\left(G_FV_{cs}^*+\frac{G_V}{\sqrt{2}} \right)+ \frac{\sqrt{q^{2}}G_{S}}{\sqrt{2}(m_{c}-m_{s})} \right|^{2} \nonumber
\right. \\
&+&\left. \mathrm{Re}\left[\left(G_FV_{cs}^*+\frac{G_V}{\sqrt{2}} \right)\frac{\mathbf{i}\ 3\sqrt{2}m_{\ell}G_T^{*}}{(m_D+m_K)}f_{T}^{K*}(q^{2})f_{+}^K(q^{2})\right]\nonumber
\right. \\
&+& \left. \frac{2\sqrt{2}(q^2+2m_{\ell}^2)}{(m_D+m_K)^2} \left|G_Tf_{T}^K(q^{2}) \right|^{2} \right\},
\end{eqnarray}
where the new form factor, $f_T^K(q^2)$, is defined for describing the contribution of the tensorial operators via\cite{Lubicz:2018rfs}
\begin{eqnarray}
&&\langle K(p_K)|\widehat{T}_{\mu\nu}|D(p_D)\rangle 
=\frac{2(p_K^\mu p_D^\nu-p_K^\nu p_D^\mu)}{m_D+m_K} f_{T}^K(q^{2}).
\label{eq:tensorf}
\end{eqnarray} 
where $\widehat{T}_{\mu\nu}$ represents the tensor operator.

To obtain constraints on $LL$- and $LR$- type fermion-leptoquark-fermion interactions from $D\rightarrow K\ell\nu_{\ell}$ decays, we re-analyze the experimental data via Eq. (\ref{eq:DKNP}) by one type new interaction at a time. The $\chi^2$ (Eq. (\ref{chitotal}) ) is re-construct to be
\begin{eqnarray}
\chi_{NP}^2 = \sum_{i=1}^{98} \sum_{j=1}^{98} \left( E_i-T_i\right)(C_{ET}^{-1})_{ij}\left( E_j-T_j\right),
\label{eq:chiNP}
\end{eqnarray}
where $E$ and $T$ are respectively the experimental value and theoretical expectation of $\Delta\Gamma$, $\mathfrak{f}$ or $f_+^{NL}$, $C_{ET}^{-1}$ is the inverse of the covariance
matrix $C_{ET}$, which is a $98\times98$ matrix. In the context of NP, $\mathfrak{f}_i^{th}$ (Eq. (\ref{eq:ffoc})) should be rewritten as
\begin{equation}
\mathfrak{f}_i^{th} = \frac{ \left[\int_{q_{i\,min}^2}^{q_{i\,max}^2}\dfrac{ \mathrm{d}\Gamma_{\mu}^{NP}/\mathrm{d}q^{2} \cdot 24\pi^3/|\mathrm{\textbf{p}}|^3/(1-m_{\mu}^2/q^2)^{2}}
	{ G_{FV}^2[V_\mu + S_\mu \left( 1+\beta q^2/\alpha \right)^2] }\mathrm{d}q^2 \right]^{\frac{1}{2}}}{f_+^K(0)\sqrt{q_{i\,max}^2-q_{i\,min}^2}},
\label{eq:ffocnp}
\end{equation}
where $G_{FV}=G_FV_{cs}^*+G_V/\sqrt{2}$ and $F_i^{th}$ (Eq. (\ref{Fith})) should be rewritten as
\begin{eqnarray}
F_i^{th} = \left[ \dfrac{\int_{q_{i\,min}^2}^{q_{i\,max}^2} \left(0.54\dfrac{ \mathrm{d}\Gamma_e^{NP}}{\mathrm{d}q^2}  + 0.46 \dfrac{ \mathrm{d}\Gamma_\mu^{NP}}{\mathrm{d}q^2} \right) \mathrm{d}q^2}{G_F^2|\mathrm{\textbf{p}_i}|^3/24\pi^3(q_{i\ max}^2-q_{i\ min}^2)} \right]^{\dfrac{1}{2}}. \nonumber \\
\label{Fithnp}
\end{eqnarray}

In our numerical calculations, a complete set of lattice calculations of $f_+^K(q^2)$ and $f_0^K(q^2)$\cite{ETM} and $f_T^K(q^2)$\cite{Lubicz:2018rfs} provided by the ETMC and $V_{cs}^*=\rvert V_{cs} \rvert=0.97351(13)$ (conventionally) obtained from the unitarity constrains\cite{PDG} are used as SM inputs. The covariance matrix $C_{ET}$ contains the correlations between the experimental measurements and the correlations between the theoretical expectations i.e.$C_{ET}=C_{exp}+C_{th}$. $C_{exp}=C_{\Delta\Gamma}\oplus C_{FOC}\oplus C_{Belle}$ where $C_{\Delta\Gamma}$, $C_{FOC}$ and $C_{Belle}$ can be obtained as the analysis in Section \ref{sec:4.2}. $C_{th}$ can be construct via the covariance among the parameters of form factors which are in the ETMC’s papers published and the uncertainty of $|V_{cs}|$. 

For $LL$ type new interactions, corresponding coefficients are real, the constraints on these coefficients, at 95\% C.L., for the case of final states with $e\nu$ pair
\begin{eqnarray}
&&\frac{\rvert \lambda_{21}^{LL} \rvert^2}{m_{S_0}^2}<5.4\times10^{-6} \ \mathrm{(GeV^{-2})}, \label{lqLL21}
\end{eqnarray}
and for the case of final states with $\mu\nu$ pair
\begin{eqnarray}
&&\frac{\rvert \lambda_{22}^{LL} \rvert^2}{m_{S_0}^2}<4.0\times10^{-6} \ \mathrm{(GeV^{-2})}. \label{lqLL22}
\end{eqnarray}

\begin{figure*}[!]
	\centering
	\vspace*{8pt}
	\subfigure[]{
		\label{fig:subfig:a} 
		\includegraphics[width=2.8in]{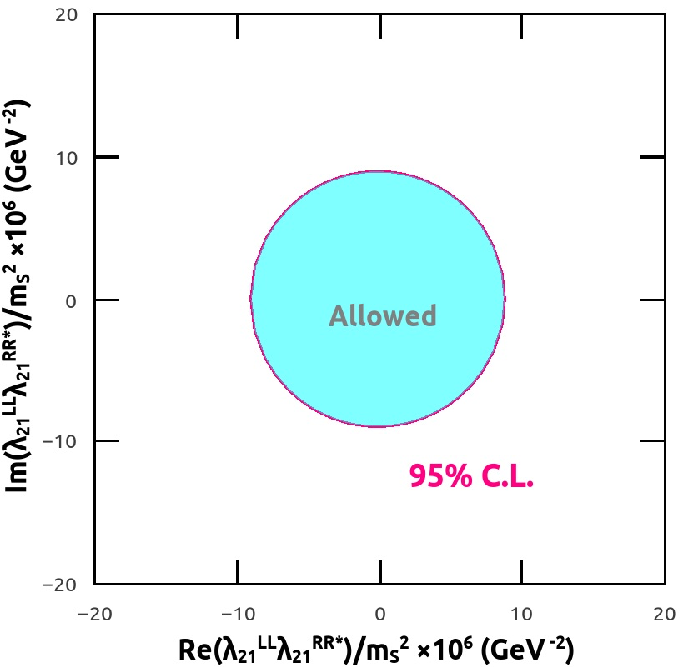}}
	\hspace{0.5in}
	\subfigure[]{
		\label{fig:subfig:b} 
		\includegraphics[width=2.8in]{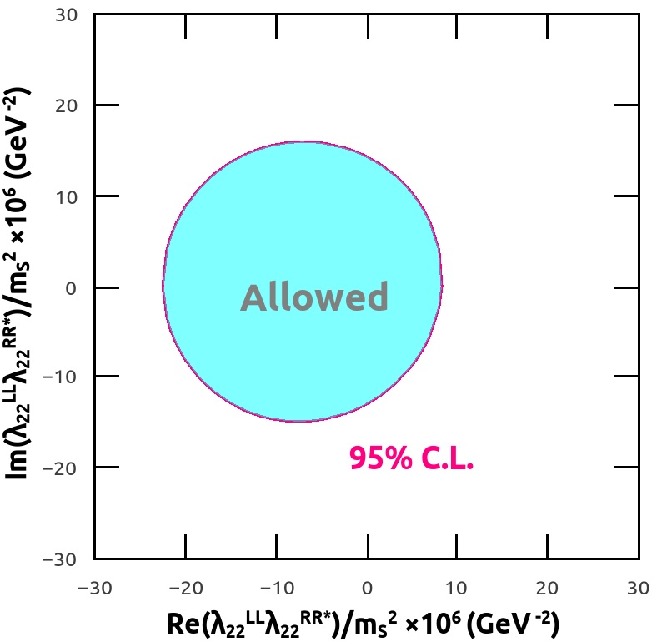}}
	\caption{Allowed regions (cyan) of (a) $\lambda_{21}^{LL}\lambda_{21}^{RR*}/m_{S_0}^2$ for the electron case and (b) $\lambda_{22}^{LL}\lambda_{22}^{RR*}/m_{S_0}^2$ for the muon case. The outside of the pink closed curves are excluded at 95\% C.L.}
	\label{fig:lqem} 
\end{figure*}

Since the coefficients corresponding to $LR$ type new interactions are complex, the 95\% C.L. curves are placed in the real-imaginary plane as shown in Figs.\ref{fig:lqem} (a) for the electron case and (b) for the muon case.

\begin{figure}[!]
	\centerline{\includegraphics[width=3.0in]{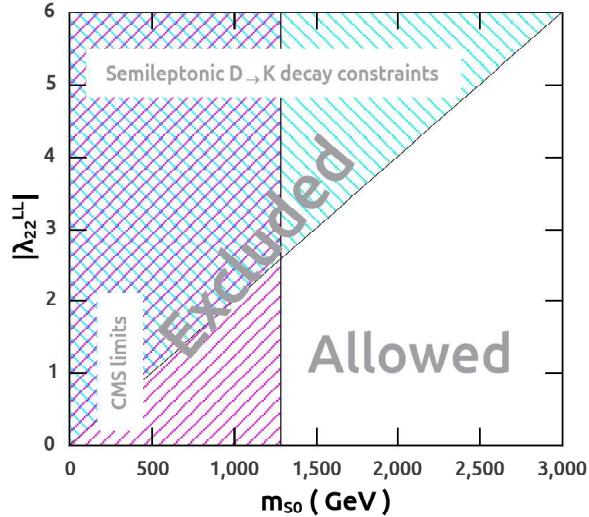}}
	\vspace*{8pt}
	\caption{Combined limits on second-generation leptoquark $S_0^L$.\protect\label{fig:lqCMS}}
\end{figure}

Recently, a search for pair production of second-genera-\ tion leptoquarks is performed by the CMS Collaboration by using 35.9 fb$^{-1}$ of data collected at $\sqrt{s}$=13 TeV in 2016 with the CMS detector at the LHC \cite{CMS:2018sgp}. By analyzing the final states with $\mu \mu jj$ and $\mu \nu jj$, they exclude second-generation leptoquarks with masses less than 1530 GeV (1285 GeV) for $\beta=1 (0.5)$ at 95\% C.L., where $\beta$ is the branching fraction of a leptoquark decaying to a charged lepton and a quark. Assuming lepton number conservation, with limits Eqs. (\ref{lqLL22}) obtained from semileptonic $D\rightarrow K$ decays in conjunction with masses limits for second-generation scalar leptoquarks obtained by the CMS Collaboration, we show the combined limits on second-generation leptoquark $S_0^L$ in Fig. \ref{fig:lqCMS}.

\section{Conclusions}
\label{sec:6}
By globally analyzing all existing measurements for $D\rightarrow K \ell^+ \nu_{\ell} (\ell=e,\ \mu)$ decays in the last 30 years, we determined both the vector and scalar form factors of $D\rightarrow K \ell^+ \nu_{\ell}$ decays from these experimental measurements. With two-parameter series expansion form factors, we obtain the product of form factor $f_+^K(0)$ and the magnitude $|V_{cs}|$ and the shape parameters of both vector and scalar form factors
\begin{eqnarray}
&&f_+^K(0)|V_{cs}|=0.7182 \pm 0.0029,\nonumber \\
&&r_{+1}=-2.16\pm0.007, \quad  r_{01}= 0.89\pm3.27. \nonumber
\end{eqnarray}
The shape parameter $r_{01}$  has a quite large uncertainty due to the small contribution of the scalar form factor to the total decay rate, 
and the precision could be improved if the experimental measurements are done with larger statistical data in future by experiments e.g.
BESIII and Belle II.

With the product $f_+^K(0)|V_{cs}|$ together with $|V_{cs}|$ obtained from unitarity constraints, we determine 
\begin{eqnarray}
f_+^K(0)=0.7377 \pm 0.003 \pm 0.000, \nonumber
\end{eqnarray}
which is consistent within error with the lattice calculations, and presents a good consistency with the previous fitting result, but with higher precision.

With the product $f_+^K(0)|V_{cs}|$ in conjunction with the average of form factor $f_+^K(0)$ from lattice calculations, the magnitude of CKM matrix element $V_{cs}$ can be extracted 
\begin{eqnarray}
\rvert V_{cs} \rvert^{D\rightarrow Kl\nu_{l}}=0.964\pm0.004\pm0.019, \nonumber
\end{eqnarray}
where the second error is from the lattice calculated form factor which is 5 times larger than the first error which is from experiments. The determined magnitude $|V_{cs}|$ presents a good consistency within error with the one from SM global fit. Then factoring in $\rvert V_{cs} \rvert^{D_s\rightarrow l\nu_{l}}=1.008\pm0.021$ determined from leptonic $D_s$ decay \cite{PDG}, the magnitude of the CKM matrix element $V_{cs}$ is determined to be 
\begin{eqnarray}
\rvert V_{cs} \rvert=0.985\pm0.014, \nonumber
\end{eqnarray}
which is in good agreement within error with the average value $\rvert V_{cs} \rvert=0.995\pm0.016$ from PDG'2016 \cite{PDG}.

We re-analyze these experimental measurements in the context of new physics. Taking the form factors determined from LQCD and $|V_{cs}|$ from unitarity constraints as input parameters, we constrain leptoquark $S_0^{-1/3}$ from $D\rightarrow K l \nu_{l}$ at 95\% C.L.. The second-generation leptoquark $S_0^L$ and relevant Yukawa couplings are constraint as
\begin{eqnarray}
&&\frac{\rvert \lambda_{22}^{LL} \rvert^2}{m_{S_0}^2}<4.0\times10^{-6} \ \mathrm{(GeV^{-2})}, \nonumber
\end{eqnarray}
Considering recent mass constraints for second-generation leptoquarks obtained by the CMS Collaboration, we give a combined limits on second-generation leptoquark $S_0^L$.

\section*{Acknowledgement}
This work was supported in part by the National Natural Science Foundation of China under Grants No.11275088 and 11747318.

\end{document}